\documentclass[aps,prc,reprint,floatfix,superscriptaddress]{revtex4-1}
\usepackage{amssymb}
\usepackage{graphicx,color}
\usepackage{latexsym}
\usepackage{array, longtable}
\usepackage{amsmath, amssymb}
\usepackage{bm}
\usepackage{docs}
\expandafter\ifx\csname package@font\endcsname\relax\else
 \expandafter\expandafter
 \expandafter\usepackage
 \expandafter\expandafter
 \expandafter{\csname package@font\endcsname}%
\fi

\begin{document}


\title{Missing-mass spectroscopy with the $^{6}$Li($\pi^{-}, K^{+}$)X reaction to search for $^{6}_{\Lambda}$H}


\author{R.~Honda}
\email[]{honda@km.phys.sci.osaka-u.ac.jp}
\affiliation{Department of Physics, Osaka University, Toyonaka, Osaka 560-0043, Japan}

\author{M.~Agnello}
\affiliation{Dipartimento di Scienza Applicata e Tecnologia, Politecnico di Torino, 1-10129 Torino, Italy}
\affiliation{INFN - Istituto Nazionale di Fisica Nucleare, Sezione di Torino, 1-10125 Torino, Italy}

\author{J.~K.~Ahn}
\affiliation{Department of Physics, Korea University, Seoul 02841, Republic of Korea}

\author{S.~Ajimura}
\affiliation{Research Center for Nuclear Physics (RCNP), 10-1 Mihogaoka, Ibaraki, Osaka 567-0047, Japan}

\author{Y.~Akazawa}
\affiliation{Department of Physics, Tohoku University, Sendai, Miyagi 980-8578, Japan}

\author{N.~Amano}
\affiliation{Department of Physics, Kyoto University, Kyoto 606-8502, Japan}

\author{K.~Aoki}
\affiliation{High Energy Accelerator Research Organization (KEK), Tsukuba 305-0801, Japan}

\author{H.~C.~Bhang}
\affiliation{Department of Physics and Astronomy, Seoul National University, Seoul 151-747, Republic of Korea}

\author{N.~Chiga}
\affiliation{Department of Physics, Tohoku University, Sendai, Miyagi 980-8578, Japan}

\author{M.~Endo}
\affiliation{Department of Physics, Osaka University, Toyonaka, Osaka 560-0043, Japan}

\author{P.~Evtoukhovitch}
\affiliation{Joint Institute for Nuclear Research, Dubna, Moscow Region 141980, Russia}

\author{A.~Feliciello}
\affiliation{INFN - Istituto Nazionale di Fisica Nucleare, Sezione di Torino, 1-10125 Torino, Italy}

\author{H.~Fujioka}
\affiliation{Department of Physics, Kyoto University, Kyoto 606-8502, Japan}

\author{T.~Fukuda}
\affiliation{Department of Engineering Science, Osaka Electro-Communication University, Neyagawa, Osaka 572-8530, Japan}

\author{S.~Hasegawa}
\affiliation{Japan Atomic Energy Agency (JAEA), Tokai, Ibaraki 319-1195, Japan}

\author{S.~H.~Hayakawa}
\affiliation{Department of Physics, Osaka University, Toyonaka, Osaka 560-0043, Japan}

\author{K.~Hosomi}
\affiliation{Japan Atomic Energy Agency (JAEA), Tokai, Ibaraki 319-1195, Japan}

\author{S.~H.~Hwang}
\affiliation{Korea Research Institute of Standards and Science (KRISS), Daejeon 34113, Republic of Korea}

\author{Y.~Ichikawa}
\affiliation{Japan Atomic Energy Agency (JAEA), Tokai, Ibaraki 319-1195, Japan}

\author{Y.~Igarashi}
\affiliation{High Energy Accelerator Research Organization (KEK), Tsukuba 305-0801, Japan}

\author{K.~Imai}
\affiliation{Japan Atomic Energy Agency (JAEA), Tokai, Ibaraki 319-1195, Japan}

\author{N.~Ishibashi}
\affiliation{Department of Physics, Osaka University, Toyonaka, Osaka 560-0043, Japan}

\author{R.~Iwasaki}
\affiliation{High Energy Accelerator Research Organization (KEK), Tsukuba 305-0801, Japan}

\author{C.~W.~Joo}
\affiliation{Depart of Physics and Astronomy, Seoul National University, Seoul 151-747, Republic of Korea}

\author{R.~Kiuchi}
\affiliation{Department of Physics, Tokai University, Hiratsuka, Kanagawa, 259-1292, Japan}

\author{J.~K.~Lee}
\affiliation{Department of Physics, Pusan National University, Busan 609-735, Republic of Korea}

\author{J.~Y.~Lee}
\affiliation{Depart of Physics and Astronomy, Seoul National University, Seoul 151-747, Republic of Korea}

\author{K.~Matsuda}
\affiliation{Department of Physics, Osaka University, Toyonaka, Osaka 560-0043, Japan}

\author{Y.~Matsumoto}
\affiliation{Department of Physics, Tohoku University, Sendai, Miyagi 980-8578, Japan}

\author{K.~Matsuoka}
\affiliation{Department of Physics, Osaka University, Toyonaka, Osaka 560-0043, Japan}

\author{K.~Miwa}
\affiliation{Department of Physics, Tohoku University, Sendai, Miyagi 980-8578, Japan}

\author{Y.~Mizoi}
\affiliation{Department of Engineering Science, Osaka Electro-Communication University, Neyagawa, Osaka 572-8530, Japan}

\author{M.~Moritsu}
\affiliation{Department of Physics, Osaka University, Toyonaka, Osaka 560-0043, Japan}

\author{T.~Nagae}
\affiliation{Department of Physics, Kyoto University, Kyoto 606-8502, Japan}

\author{S.~Nagamiya}
\affiliation{Japan Atomic Energy Agency (JAEA), Tokai, Ibaraki 319-1195, Japan}

\author{M.~Nakagawa}
\affiliation{Department of Physics, Osaka University, Toyonaka, Osaka 560-0043, Japan}

\author{M.~Naruki}
\affiliation{Department of Physics, Kyoto University, Kyoto 606-8502, Japan}

\author{H.~Noumi}
\affiliation{Research Center for Nuclear Physics (RCNP), 10-1 Mihogaoka, Ibaraki, Osaka 567-0047, Japan}

\author{R.~Ota}
\affiliation{Department of Physics, Osaka University, Toyonaka, Osaka 560-0043, Japan}

\author{B.~J.~Roy}
\affiliation{Nuclear Physics Division, Bhabha Atomic Research Center (BARC), Trombay, Mumbai 400 085, India}

\author{P.~K.~Saha}
\affiliation{Japan Atomic Energy Agency (JAEA), Tokai, Ibaraki 319-1195, Japan}

\author{A.~Sakaguchi}
\affiliation{Department of Physics, Osaka University, Toyonaka, Osaka 560-0043, Japan}

\author{H.~Sako}
\affiliation{Japan Atomic Energy Agency (JAEA), Tokai, Ibaraki 319-1195, Japan}

\author{C.~Samanta}
\affiliation{Department of Physics and Astronomy, Virginia Military Institute, Lexington, VA 24450, USA}

\author{V.~Samoilov}
\affiliation{Joint institute for Nuclear Research, Dubna, Moscow Region 141980, Russia}

\author{Y.~Sasaki}
\affiliation{Department of Physics, Tohoku University, Sendai, Miyagi 980-8578, Japan}

\author{S.~Sato}
\affiliation{Japan Atomic Energy Agency (JAEA), Tokai, Ibaraki 319-1195, Japan}

\author{M.~Sekimoto}
\affiliation{High Energy Accelerator Research Organization (KEK), Tsukuba 305-0801, Japan}

\author{Y.~Shimizu}
\affiliation{Department of Engineering Science, Osaka Electro-Communication University, Neyagawa, Osaka 572-8530, Japan}

\author{T.~Shiozaki}
\affiliation{Department of Physics, Tohoku University, Sendai, Miyagi 980-8578, Japan}

\author{K.~Shirotori}
\affiliation{Research Center for Nuclear Physics (RCNP), 10-1 Mihogaoka, Ibaraki, Osaka 567-0047, Japan}

\author{T.~Soyama}
\affiliation{Department of Physics, Osaka University, Toyonaka, Osaka 560-0043, Japan}

\author{H.~Sugimura}
\affiliation{Japan Atomic Energy Agency (JAEA), Tokai, Ibaraki 319-1195, Japan}

\author{T.~Takahashi}
\affiliation{High Energy Accelerator Research Organization (KEK), Tsukuba 305-0801, Japan}

\author{T.~N.~Takahashi}
\affiliation{Research Center for Nuclear Physics (RCNP), 10-1 Mihogaoka, Ibaraki, Osaka 567-0047, Japan}
\affiliation{RIKEN, 2-1 Hirosawa, Wako, Saitama, 351-0198, Japan}

\author{H.~Tamura}
\affiliation{Department of Physics, Tohoku University, Sendai, Miyagi 980-8578, Japan}

\author{K.~Tanabe}
\affiliation{Department of Physics, Tohoku University, Sendai, Miyagi 980-8578, Japan}

\author{T.~Tanaka}
\affiliation{Department of Physics, Osaka University, Toyonaka, Osaka 560-0043, Japan}

\author{K.~Tanida}
\affiliation{Japan Atomic Energy Agency (JAEA), Tokai, Ibaraki 319-1195, Japan}

\author{A.~O.~Tokiyasu}
\affiliation{Research Center For Electron Photon Science (ELPH), Sendai, Miyagi 982-0826, Japan}

\author{Z.~Tsamalaidze}
\affiliation{Joint institute for Nuclear Research, Dubna, Moscow Region 141980, Russia}

\author{M.~Ukai}
\affiliation{High Energy Accelerator Research Organization (KEK), Tsukuba 305-0801, Japan}

\author{T.~O.~Yamamoto}
\affiliation{High Energy Accelerator Research Organization (KEK), Tsukuba 305-0801, Japan}

\author{Y.~Yamamoto}
\affiliation{Research Center for Nuclear Physics (RCNP), 10-1 Mihogaoka, Ibaraki, Osaka 567-0047, Japan}

\author{S.~B.~Yang}
\affiliation{Depart of Physics and Astronomy, Seoul National University, Seoul 151-747, Republic of Korea}

\author{K.~Yoshida}
\affiliation{Department of Physics, Osaka University, Toyonaka, Osaka 560-0043, Japan}

\collaboration{J-PARC E10 Collaboration}

\date{\today}

\begin{abstract}
  We searched for the bound state of the neutron-rich $\Lambda$-hypernucleus $^{6}_{\Lambda}$H, using the $^{6}$Li($\pi^{-}, K^{+}$)X double charge-exchange reaction at a $\pi^{-}$ beam momentum of 1.2 GeV/$c$ at J-PARC.
  A total of $1.4 \times 10^{12}\;\pi^{-}$ was driven onto a $^{6}$Li target of 3.5-$\mathrm{g/cm^{2}}$ thickness.
  No event was observed below the bound threshold, i.e., the mass of $^{4}_{\Lambda}{\mathrm H}+2n$, in the missing-mass spectrum of the $^{6}$Li($\pi^{-}, K^{+}$)X reaction in the $2^{\circ} < \theta_{\pi K} < 20^{\circ}$ angular range.
  Furthermore, no event was found up to 2.8 MeV/$c^{2}$ above the bound threshold. 
  We obtained the the double-differential cross section spectra of the $^{6}$Li($\pi^{-}, K^{+}$)X reaction in the angular range of $2^{\circ} < \theta_{\pi K} < 14^{\circ}$.
  An upper limit of 0.56 nb/sr (90\% C.L.) was obtained for the production cross section of the $^{6}_{\Lambda}$H hypernucleus bound state.
  In addition, not only the bound state region, but also the $\Lambda$ continuum region and part of the $\Sigma^{-}$ quasi-free production region of the $^{6}$Li($\pi^{-}, K^{+}$) reaction, were obtained with high statistics.
  The present missing-mass spectrum will facilitate the investigation of the $\Sigma^{-}$-nucleus optical potential for $\Sigma^{-}$-$^{5}$He through spectrum shape analysis.
\end{abstract}


\maketitle

\section{Introduction}
\label{sec_introduction}
The investigation of neutron-rich $\Lambda$-hypernuclei is an important subject in hypernuclear physics, because it facilitates the understanding of both the structure of neutron-rich nuclei flavored by the $\Lambda$ hyperon and the $\Lambda N$-$\Sigma N$ mixing effect in the hyperon-nucleon ($YN$) interaction.
When a $\Lambda$ particle is introduced in a nucleus, it can deeply penetrate its interior and it can form a $\Lambda$-hypernucleus because the $\Lambda$ is not affected by the Pauli blocking from the other nucleons.
A $\Lambda$ particle attracts surrounding nucleons, i.e., through the glue-like role of the $\Lambda$.
Then, the system is bound more deeply.
This glue-like role stabilizes several unstable nuclear systems, like $^{9}_{\Lambda}$Be \cite{Be9L} and $^{6}_{\Lambda}$He \cite{Emulsion}.
Actually, the ground states of $^{8}$Be and $^{5}$He are particle unstable.
Thus, the results of experimental studies on the masses of the $\Lambda$-hypernuclei are incorporated into theoretical treatment of the particle-unstable nuclei \cite{FBHiyama}.
In particular, the $\Lambda N$ interaction could change the nucleus structure such as the neutron halo on the neutron-rich side of the nuclear chart.
Another important aspect of the neutron-rich $\Lambda$-hypernuclei is a possible enhancement of the $\Sigma$ hyperon mixing, which is known as the $\Lambda$-$\Sigma$ coupling \cite{Gibson_LS}.
It is well known that the $\Lambda$ and the $\Sigma$ are not mixed in free space, because of the difference in their isospins.
However, in nuclei, the $\Sigma$ may appear in the intermediate state of $\Lambda N$, $\Lambda NN$ interactions, and so on.
Consequently, $\Sigma$ admixture with the $\Lambda$-hypernuclear state is allowed with no excitation of the core nuclei in the case of non-zero isospin.
Akaishi \textit{et al.} have suggested that this feature of the $\Lambda$-$\Sigma$ coupling is essential for explaining the energy levels of the $\mathrm{A = 4}$ $\Lambda$-hypernuclei \cite{LSAkaishi}.
In neutron-rich $\Lambda$-hypernuclei, the $\Lambda N$-$\Sigma N$ mixing ($\Lambda$-$\Sigma$ coupling) effect is presumably enhanced owing to the large isospin value of the core nucleus, which can act as a buffer to the isospin for the $\Sigma$ mixing.

To investigate the $YN$ interactions, not only the ground state, but also the exited states must be studied experimentally.
For this purpose, counter experiments exploiting the missing-mass spectroscopy via the ($K, \pi$) and ($\pi, K$) reactions represent the most straightforward approach.
The non-charge-exchange processes (NCX), such as the ($K^{-}_{\mathrm{stopped}}, \pi^{-}$), the in-flight ($K^{-}, \pi^{-}$), and the ($\pi^{+}, K^{+}$) reactions, have primarily been employed for the production of $\Lambda$-hypernuclei \cite{hyp1, hyp2, hyp3, hyp4, hyp5}.
In these reactions, one neutron is converted to a $\Lambda$.
On the other hand, as the double charge-exchange (DCX) reactions, such as the ($K^{-}, \pi^{+}$) and the ($\pi^{-}, K^{+}$) reactions, convert two protons into one $\Lambda$ and into one neutron, neutron-rich $\Lambda$-hypernuclei can be produced \cite{NRichHyp_Majling}.
In particular, DCX reactions involving light target nuclei produces neutron-rich $\Lambda$-hypernuclei with a quite high $N/Z$ ratio.

In the KEK-PS E521 experiment, the production of the neutron-rich $\Lambda$-hypernucleus $^{10}_{\Lambda}$Li was successfully achieved for the first time via the $^{10}$B($\pi^{-}, K^{+}$) reaction \cite{E521}.
Events below the $\Lambda$ binding threshold were clearly observed.
However, the binding energy of the ground state was not determined, because a peak structure was not clearly observed.
The obtained result also put in evidence that the cross section of the $^{10}$B($\pi^{-}, K^{+}$)$^{10}_{\Lambda}$Li reaction was significantly smaller than that of the ($\pi^{+}, K^{+}$) reaction.
The integrated cross section in the $\Lambda$ bound region was 11.3$\:\pm\:$1.9 nb/sr \cite{E521} at a beam momentum of 1.2 GeV/$c$.
This cross section was roughly three orders of magnitude lower than the typical production cross section of $\Lambda$-hypernuclei via the ($\pi^{+}, K^{+}$) reaction.
The FINUDA Collaboration performed an experiment to produce the neutron-rich $\Lambda$-hypernuclei $^{6}_{\Lambda}$H and $^{7}_{\Lambda}$H via the $^{6,7}$Li($K^{-}_{\mathrm{stopped}}, \pi^{+}$) reaction \cite{FINUDA1}.
As a first step, they set a production rate upper limit \cite{FINUDA1}. 
More recently, they reported three candidate events for the $^{6}_{\Lambda}$H production and decay \cite{FINUDA2}. 
The production rate of the neutron-rich $\Lambda$-hypernucleus via the DCX reaction was also small in the FINUDA experiment. 
A $^{6}_{\Lambda}$H production rate of (5.9$\:\pm\:$4.0)$\times 10^{-6}/K^{-}_{stop}$ was obtained \cite{FINUDA2}, which can be compared with that of the $^{12}_{\Lambda}$C and $^{4}_{\Lambda}$He productions via the NCX reaction, ($K^{-}_{\mathrm{stopped}}, \pi^{-}$) \cite{FINUDA_NCX, Kpi_4LHe}.

The FINUDA result has triggered extensive discussions regarding the existence of the $^{6}_{\Lambda}$H bound state.
The binding energy of the $^{6}_{\Lambda}$H hypernucleus was first predicted theoretically by Dalitz and Levi~Setti \cite{Dalitz_H6L}. 
It should be noted that at that time $^{5}$H was believed to be bound.
Today, $^{6}_{\Lambda}$H is known to be a quite exotic system, not only because of the high $N/Z$ ratio, but also because the core nucleus $^{5}$H was observed as a resonance \cite{H5_1}.
To date, several theoretical works have been conducted regarding the $\mathrm{B_{\Lambda}}$ of the $^{6}_{\Lambda}$H ground state.
Akaishi and Yamazaki have suggested that this system may be deeply bound by $\mathrm{B_{\Lambda}} = 5.8$ MeV with respect to the $^{5}{\mathrm{H}} + \Lambda$ system, owing to the additional attraction due to the coherent $\Lambda$-$\Sigma$ coupling in the neutron-rich environment \cite{Akaishi_H6L}.
A rather smaller $\mathrm{B_{\Lambda}} = 3.83 \pm 0.08 \pm 0.22$ MeV was predicted by Gal and Millener, based on the shell-model calculation \cite{Gal_H6L}.
On the other hand, Hiyama \textit{et~al.} \cite{Hiyama_H6L} have noted that this system is 0.87 MeV unbound above the $^{4}_{\Lambda}{\mathrm{H}}+2n$ threshold within the framework of the four-body cluster model, tuned using interactions that reproduced the experimentally observed resonance energy and the width of $^{5}$H \cite{H5_1}.
The mass of the $^{4}_{\Lambda}{\mathrm{H}}+2n$ system is roughly 3.7 MeV/$c^{2}$ smaller than that of the $^{5}{\mathrm{H}} + \Lambda$ system.
In that study, the broad spatial distribution of $^{5}$H contradicted the existence of the bound state of $^{6}_{\Lambda}$H, because the wave function overlap between the $\Lambda$ and nucleus was small.
However, those researchers also suggested a possible binding by adjusting the $tnn$ three-body force within the uncertainty of the experimental result for $^{5}$H.
The existence of the bound state of $^{6}_{\Lambda}$H and its binding energy are closely related to not only the $YN$ interaction, but also the structure of the unstable core nucleus $^{5}$H.

The neutron-rich $\Lambda$-hypernuclei have also been studied in terms of the reaction mechanism.
In the production of neutron-rich $\Lambda$-hypernuclei via the ($\pi^{-}, K^{+}$) reaction, the missing-mass spectrum shape provides information about the production mechanism.
Two nucleons must be involved in order to produce a $\Lambda$ particle via the DCX reaction, because of the charge conservation.
Then, two different processes are considered for the $\Lambda$ production, namely, the two-step and one-step processes.
The two-step process consists of a series of single-charge exchange reactions, i.e., $\pi^{-} + pp \rightarrow \pi^{0} + (pn) \rightarrow K^{+} + \Lambda n$ or $\pi^{-} + pp \rightarrow K^{0} + (\Lambda p) \rightarrow K^{+} + \Lambda n$, where parentheses indicate off-shell kinematics.
On the other hand, the neutron-rich $\Lambda$-hypernuclei is directly produced through an $\Sigma$ admixed in the ground state of the $\Lambda$-hypernuclei in the one-step process.
The one-step process is expressed as $\pi^{-} + pp \rightarrow K^{+} + (\Sigma^{-} p) \leftrightarrow K^{+} + \Lambda n$.
Harada $et~al.$ have analyzed the KEK-PS E521 data theoretically using the distorted-wave impulse approximation (DWIA) calculation and they suggested that, in that case, the one-step process was dominant while the contribution from the two-step process was quite small \cite{Harada_E521}.
The production cross section of the bound states depends on the admixture probability of the virtual $\Sigma^{-}$ in the $\Lambda$-hypernuclear states.
The most reasonable admixture probability value reproducing the experimental result is 0.58\% \cite{Harada_E521} in the low-lying state of $^{10}_{\Lambda}$Li.
In the case of the one-step process, the production cross section of the $\Lambda$-hypernuclei is affected not only by the $\Lambda$-nucleus optical potential and the $\Sigma^{-}$ admixture probablity, but also by the $\Sigma^{-}$-nucuelus optical potential.
As the virtual $\Sigma$ state is a doorway for the formation of the $\Lambda$ bound state, an overlap between the wave functions of the core nucleus and of the $\Sigma$ appeares in the cross section calculation.
Therefore, determination of the $\Sigma$-nucleus optical potential is essential in order to examine the global spectrum shape of the ($\pi^{-}, K^{+}$) reaction, together with the bound states, via the one-step process. 

Historically, the spectrum shape analysis has played an important role in extracting the $\Sigma$-nucleus optical potential, for instance, in the analysis performed with the ($\pi^{-},K^{+}$) spectra obtained in the KEK-PS E438 experiment \cite{E438}.
The strengths of the real and imaginary parts of the potential were estimated based on the spectral shape in the $\Sigma^{-}$ quasi-free production region above the $\Sigma$ binding threshold and the $\Lambda$ continuum below the threshold, respectively.
Hence, experimental data from the $^{6}_{\Lambda}$H bound region to the $\Sigma^{-}$ quasi-free production region were essential for determining the $\Lambda$-nucleus and $\Sigma$-nucleus optical potentials and the $\Sigma^{-}$ admixture probablity simultaneously through spectrum fitting with the DWIA calculation.

The missing-mass spectrum of the $^{6}$Li($\pi^{-}, K^{+}$)X reaction with high statistics and in a wide missing-mass range is required in order to confirm the existence of the $^{6}_{\Lambda}$H hypernucleus through the theoretical investigation.
For the physics-based motivations introduced above, the first stage of the Japan Proton Accelerator Research Complex (J-PARC) E10 experiment, which aimed to search for the $^{6}_{\Lambda}$H hypernucleus, was conducted.
The result of the first analysis has already been reported, that is, three events were observed below the $^{4}_{\Lambda}{\mathrm{H}} + 2n$ threshold \cite{Sugimura}.
However, we were unable to conclude whether events below the $^{4}_{\Lambda}{\mathrm{H}} + 2n$ threshold corresponded to $^{6}_{\Lambda}$H signals.
The expected number of background events was also 2.1 in the missing-mass window, and then an upper limit of 1.2 nb/sr with 90\% of confidence level (C.L.) was obtained for the $^{6}_{\Lambda}$H production cross section \cite{Sugimura}.
In order to draw more definitive conclusions, the background reduction method, the missing-mass resolution, and the analysis efficiencies were improved.
In this paper, we describe the results of the updated analysis.

\section{Experiment}
\begin{figure}[htb]
  \includegraphics[width=6.0cm,clip]{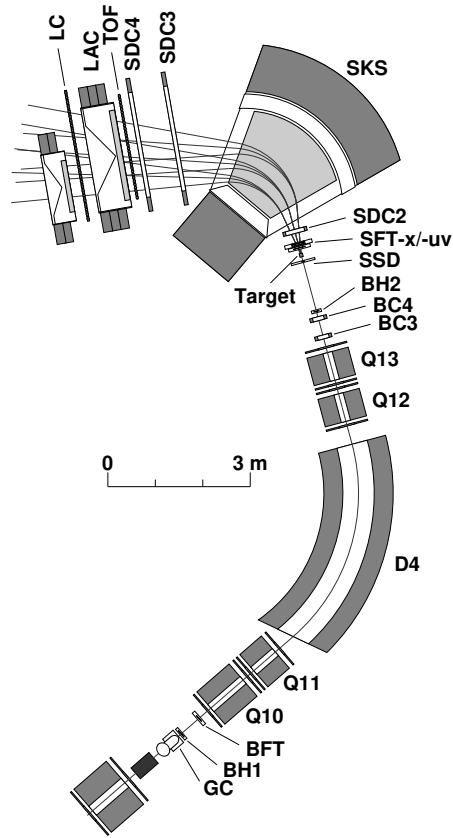}
  \caption{Schematic view of the experimental setup.}
  \label{fig:k18}
\end{figure}

The J-PARC E10 experiment was conceived to observe the production of the $^{6}_{\Lambda}$H hypernucleus via the $^{6}$Li($\pi^{-}, K^{+}$)X reaction using missing-mass spectroscopy.
The experiment was performed at the K1.8 beam line of the J-PARC Hadron Experimental Facility.
An enriched $^{6}$Li target of 3.5-$\mathrm{g/cm^{2}}$ thickness was irradiated with $1.4 \times 10^{12}$ $\pi^{-}$ beams at a beam momentum of 1.2 GeV/\textit{c}.
The experimental setup is shown in Fig.~\ref{fig:k18}.
The $\pi^{-}$ beam was analyzed by the magnetic spectrometer placed in the K1.8 experimental area, the K1.8 beam line spectrometer \cite{Takahashi_K18}, while the momentum of the outgoing $K^{+}$ was measured by the superconducting kaon spectrometer (SKS) complex \cite{Takahashi_K18, Fukuda_SKS}.
In this section, the details of the experimental apparatus are described.


\subsection{Experimental apparatus}
\subsubsection{K1.8 beam line}
The 30-GeV primary proton beam was extracted from the J-PARC main ring (MR) into the Hadron Experimental Facility using the slow extraction method.
The duration of beam extraction cycle was 6 s and the beam spill lasted typically 2 s.
The primary beam bombarded the production target, an Au rod (6 mm $\phi$ and 60-mm length). 
Charged particles generated in the production target were transported along the K1.8 beam line.
The K1.8 beam line was a general-purpose beam line equipped with double electrostatic separators (ESS) to transport well-mass-separated secondary hadron beams with a momentum up to 2.0 GeV/$c$.

\subsubsection{K1.8 beam line spectrometer}
The beam momentum was analyzed and the beam was focused on the experimental target by the K1.8 beam line spectrometer.
The typical beam size on the experimental target was $56^{\mathrm{H}} \times 28^{\mathrm{V}} \mathrm{mm^{2}}$ full width at tenth maximum (FWTM).

The spectrometer consisted of several analyzer magnets in a QQDQQ configuration, tracking detectors, and trigger counters.
The incoming particle hit coordinates were measured by the beam line fiber tracker (BFT) and multi-wire drift chambers (BC3 and 4).
The different types of particle present in the incident beam were identified by the trigger counters, gas \v{C}erenkov counter (GC), and beam hodoscopes (BH1 and 2).

GC consisted of a radiator filled with isobutyl alcohol gas (0.15 MPa), and it was used to estimate the degree of electron contamination in the $\pi^{-}$ beam.  
GC was placed at the most upstream part of the spectrometer.
Immediately downstream of GC, BH1 was installed; this detector consisted of a plastic scintillation counter segmented into 11 modules.
A second hodoscope (BH2), again a plastic scintillation counter with 8 segments, was placed at the exit of the QQDQQ magnets.
The typical timing resolution of the beam time-of-flight between BH1 and 2 was roughly 300 ps (rms) under the high-beam-intensity condition, with $12 \times 10^{6}$ pions per spill.
As the $K^{-}$ and antiproton contamination were negligibly small, because of the double ESS system, the events due to incoming $\pi^{-}$ were identified from the BH2 hit signal only.
GC and BH1 did not participate in the trigger to avoid high rejection rate and counting loss under the high-counting-rate condition, respectively.

BFT measured the horizontal coordinate of the beam particles \cite{BFT} at the entrance of the QQDQQ magnets, being composed of 1-mm diameter scintillating fibers and having a sensitive area of $160^{\mathrm{H}} \times 80^{\mathrm{V}}$ ${\mathrm{mm^{2}}}$.
BFT consisted of two scintillating fiber layers staggered by 0.5 mm and placed in contact with each other to reduce the insensitive regions.
The scintillation light was detected by pixelated photon detectors (Hamamatsu multi-pixel photon counter, MPPC) connected fiber-by-fiber with an Extended Analogue SiPM Integrated ReadOut Chip (EASIROC) system \cite{EASIROC}.
The timing resolution was the key factor in suppressing accidental hits under the high-rate condition.
BFT was able to identify particles with a timing resolution of 0.68 ns (rms) and simultaneously measure hit coordinates with a resolution of 190 $\mu$m (rms) \cite{BFT}.
The beam trajectories at the analyzer magnet exit were measured by two multi-wire drift chambers (BC3 and 4).
The anode wire spacing of BC3 and 4 was 3 mm and the sensitive area was $192^{\mathrm{H}} \times 100^{\mathrm{V}}$ $\mathrm{mm^{2}}$ \cite{Takahashi_K18}.
BC3 and 4 were installed between the Q13 magnet and BH2, and had the same structure with six planes ($xx', uu', vv'$).
The $xx'$, $uu'$, and $vv'$ pairs were in the pair plane configuration and the $u$ and $v$ wires were tilted by $15^{\circ}$ and $-15^{\circ}$ with respect to the vertical ($x$) wires, respectively.
The spatial resolution in each plane was roughly 200 $\mu$m (rms).
A gas mixture of Ar (76\%), iso-$\mathrm{C_{4}H_{10}}$ (20\%), and methylal (4\%) was used.
The beam momentum was reconstructed using the spatial information from BFT, BC3, and BC4 and the third-order transfer-matrix calculated by \textsc{orbit} \cite{ORBIT}.
The magnetic field of the dipole magnet was continuously monitored using a high-resolution Hall probe and the field fluctuation was less than 0.01\%.

Two silicon-strip detectors (SSD) with an 80-$\mu$m strip pitch and a sensitive area of $62^{\mathrm{H}} \times 61.6^{\mathrm{V}}$ $\mathrm{mm^{2}}$ were installed in front of the target as a vertex detectors.
Both the horizontal and vertical coordinates of the beam particles were measured by the SSDs.
The beam trajectories obtained by BC3 and 4 were corrected by the SSDs in order to improve the vertex resolution.
A helium bag was installed between BH2 and the SSDs to reduce the multiple scattering effect of the air.

\subsubsection{SKS complex}
The SKS complex was comprised of a superconducting dipole magnet, four tracking detectors, and trigger counters.
This system was originally used for hypernuclear spectroscopy at the KEK-PS K6 beam line and later moved to the K1.8 beam line at J-PARC.
A large effective solid angle of roughly 100 msr was realized owing to a wide aperture.
The SKS magnet was excited to 2.16 T in the present experiment.
The spectrometer complex measured the scattered particles in the 0.7--1.1 GeV/\textit{c} momentum range.
The momentum of a scattered $K^{+}$, when the expected ground state of $^{6}_{\Lambda}$H was produced, fell at the center of the momentum acceptance,  having a value of approximately 0.9 GeV/\textit{c}.

At the entrance of the SKS magnet, a scattered kaon fiber tracker (SFT) and a multi-wire drift chamber (SDC2) were installed.
SFT was a tracking detector placed immediately behind the target; the detector had high-rate capability.
The detector was installed at this position because the single rate per unit area was high as the result of the beam focusing around the target.
SFT consisted of three independent sensitive planes ($x$,$u$, and $v$).
The $u$ and $v$ fibers were tilted by $\mathrm{45^{\circ}}$ and $\mathrm{-45^{\circ}}$ with respect to the vertical ($x$) fibers, respectively.
SFT had a sensitive area of $256^{\mathrm{H}} \times 160^{\mathrm{V}}$ $\mathrm{mm^{2}}$.
The structure of the $x$ plane was identical to that of BFT, that is, two layers consisting of 1-mm diameter scintillating fibers were in contact with each other and one layer was staggered by 0.5 mm with respect to the other one.
The $u$ and $v$ planes were created by scintillating fibers with a diameter of 0.5 mm to maintain the lowest possible energy-loss straggling and multiple scattering.
In the $u$ and $v$ planes, the scintillation light from three adjacent fibers was detected by one MPPC in order to reduce the total number of readout channels. 
The timing resolutions of the $x$ plane and of the $u$ and $v$ planes were 0.8 and 1.3 ns (rms), respectively.
The SFT spatial resolutions were 190 $\mu$m (rms) for the $x$ plane and 270 $\mu$m (rms) for the $u$ and $v$ planes.
SDC2 was a multi-wire drift chamber consisting of six planes ($xx', uu', vv'$) with an anode wire spacing of 5 mm, and featuring pair plane configuration \cite{Takahashi_K18}.
Its sensitive area was $400^{\mathrm{H}} \times 150^{\mathrm{V}}$ $\mathrm{mm^{2}}$.
The tilt angles of the $u$ and $v$ wires were $\pm 15^{\circ}$ with respect to the $x$ ones.
The spatial resolution of each plane in SDC2 was roughly 200 $\mu$m (rms).
SDC2 was filled with the same gas mixture used for BC3 and 4.

Two large multi-wire drift chambers (SDC3 and 4) with a sensitive area of $2140^{\mathrm{H}} \times 1140^{\mathrm{V}}$ $\mathrm{mm^{2}}$ were used as tracking detectors at the exit of the SKS magnet.
SDC3 and 4 had identical structures and six planes each ($xuvxuv$).
The cell size and spatial resolution were 10 mm and roughly 300 $\mu$m (rms), respectively.
Helium bags were installed in the pole gap of the SKS magnet and immediately behind SDC3 to reduce the multiple scattering effects.
Ar (50\%) - $\mathrm{C_{2}H_{6}}$ (50\%) gas mixture were used for SDC3 and 4.
The momenta of the scattered particles were obtained by calculating the trajectories in the SKS magnet using the 4th-order Runge-Kutta method \cite{RungeKutta}.

The time-of-flight of the scattered particles was measured by BH2 and the TOF wall located downstream of SDC4.
The TOF wall consisted of 32 plastic scintillator modules with dimensions of 1000 mm (height), 70 mm (width), and 30 mm (thickness).
The typical time-of-flight resolution for particles traveling between BH2 and the TOF wall was roughly 200 ps (rms).
A large silica-aerogel \v{C}erenkov counter (LAC) was installed behind the TOF wall for $\pi^{+}$ veto.
The refractive index of the LAC radiators was 1.05 and the LAC was sensitive to charged particles with $\beta$ higher than 0.95.
A Lucite \v{C}erenkov counter (LC) was located immediately downstream of the LAC to discriminate slow protons from $\pi^{+}$ and $K^{+}$.
This device contained acrylic radiators segmented into 28 modules with a refractive index of 1.49.
The $\beta$ threshold of the LC was 0.67.

\subsection{Trigger}
The ($\pi^{-}, K^{+}$) events were selected by the 1st- and 2nd-level triggers.
The 1st-level $\pi K$ trigger was given by the coincidence of the trigger-counter fast signals.
It was expressed as  $\mathrm{BH2 \times TOF \times \overline{LAC} \times LC}$.
The typical trigger rate was 1200 per spill at a beam intensity of $12\times10^{6}$ pions per spill.
Furthermore, the 2nd-level trigger was introduced in order to reject protons with a velocity larger than the $\beta$ threshold of LC.
The momenta of the scattered particles were strongly correlated with the hit combination between the TOF and LC segments. 
Then, the time-of-flight between BH2 and the TOF wall for protons clearly differed from that of the $K^{+}$ after selection of the hit combination.
The 2nd-level trigger system was implemented with a field-programmable gate array (FPGA) system that gathered the time-of-flight information digitized by the Fast Encoding and Readout ADC (FERA) system.
The time-of-flight range was chosen based on preset values and the 2nd-level trigger was generated.

Owing to the 2nd-level trigger, the trigger rate was reduced to roughly half of the 1st-level trigger.

\subsection{Experimental target and data summary}
The experimental target was a slice of $^{6}$Li (95.54\% enriched) with 3.5-$\mathrm{g/cm^{2}}$ thickness.
Its cross sectional size was $\mathrm{70^{H} \times 40^{V}}$ $\mathrm{mm^{2}}$.
The target was doubly packaged with transparent bags having 55-$\mu$m thickness, and filled with Ar gas to suppress chemical deterioration of the Li.
The bag had a two-layer structure made of polyethylene (35 $\mu$m) and polyvinylidene-coated polypropylene (20 $\mu$m) layers.
During the experiment, no visible chemical deterioration of the Li was observed.
In addition, graphite (3.6 $\mathrm{g/cm^{2}}$) and polyethylene (3.4 $\mathrm{g/cm^{2}}$) targets were used to obtain calibration data.
The cross sectional sizes of the graphite and polyethylene targets were $\mathrm{80^{H} \times 88^{V}}$ and $\mathrm{80^{H} \times 40^{V}}$ $\mathrm{mm^{2}}$, respectively.

The acquired data sets are summarized in Table \ref{tb:datasummary}.
The total number of $\pi^{-}$ injected onto the $^{6}$Li target was $1.4\times 10^{12}$.
The missing-mass resolution was estimated using data set B, while data sets C, D, and E were used for momentum calibration.
The energy loss in the $^{6}$Li target was estimated using data sets E and E'.

\begin{table*}[htb]
  \caption{Data summary. $N_{pion}$ indicates the total number of injected $\pi^{\pm}$ beams.}
  \begin{tabular*}{14cm}{@{\extracolsep{\fill}}ccccccc} \hline \hline
    Data                  & Momentum           & Reaction           & Target       & Intensity       & $N_{pion}$ & Tag \\ 
                          & (GeV/\textit{c})      &                    &              & ($10^{6}$/spill)&                &    \\ \hline
    $^{6}_{\Lambda}$H     & 1.2                & ($\pi^{-}, K^{+}$) & $^{6}$Li     & 12              & $1.4 \times 10^{12}$ & Set A \\
    $^{12}_{\Lambda}$C    & 1.2                & ($\pi^{+}, K^{+}$) & graphite     & 4               & $3.0 \times 10^{10}$ & Set B \\
    $\Sigma^{-}$          & 1.39               & ($\pi^{-}, K^{+}$) & polyethylene & 10              & $2.4 \times 10^{10}$ & Set C \\
    $\Sigma^{+}$          & 1.39               & ($\pi^{+}, K^{+}$) & polyethylene & 3.5             & $2.5 \times 10^{9}$  & Set D \\
    beam through          & 0.8, 0.9, 1.0, 1.2 &                    & none         &                 &                          & Set E \\
    beam through          & 0.8, 0.9, 1.0, 1.2 &                    & $^{6}$Li     &                 &                          & Set E' \\ \hline \hline
  \end{tabular*}
  \label{tb:datasummary}
\end{table*}

\section{Analysis}
The cross section of the $^{6}$Li($\pi^{-}, K^{+}$)X reaction was derived according to the following procedures.
First, the momenta of the beam and of scattered particles were reconstructed.
Then, we identified $\pi^{-}$ and $K^{+}$ and calculated the reaction vertex position in order to select the $^{6}$Li($\pi^{-}, K^{+}$)X events.
After the event selection, the reconstructed momenta were corrected according to the momentum correction function obtained from the beam through data and the $\Sigma^{\pm}$ missing-mass peaks.
Finally, we obtained the raw missing-mass spectrum of the $^{6}$Li($\pi^{-}, K^{+}$)X reaction with no efficiency correction.
In order to obtain the cross section value, we estimated efficiencies such as the analysis efficiency of each detector, the $K^{+}$ decay factor and the acceptance of the SKS complex.
In this section, we describe the details of these analysis procedures.

\subsection{Momentum reconstruction}
\subsubsection{K1.8 beam line spectrometer}
Before the momentum reconstruction, BFT hits were selected using a time gating of $\pm$3.0 ns and by checking the hit position matching between BFT and BH1.
The local straight tracks in BC3 and 4 were selected based on tracking $\chi^{2}$ and spatial alignment matching with the BH2 hit segment.
The beam momentum was reconstructed using the third-order transfer-matrix.
The kinetic property of a beam particle moving in the beam-line components was represented by the vector
\begin{equation}
  \bm{X} = \left(x, x^{'}, y, y^{'}, \delta \right).
\end{equation}
Here, $x$ and $y$ are the beam particle coordinates in the horizontal and the vertical directions, respectively, $x^{'}$ and $y^{'}$ are tangents with respect to the central trajectory, and $\delta$ is the fractional momentum deviation from the central value.
The vector at $VO$, which is the reference point at the analyzer exit, was reversely transformed by a transfer matrix to a vector at BFT.
Then, the $x$ coordinate at BFT was described using matrices labeled $R, T$, and $U$, which correspond the 1st-, the 2nd-, and the 3rd-order matrices, respectively,
\begin{eqnarray}
\label{eq:transferMT}
x(BFT) =&& \sum_{j}R_{j}X_j(VO)\nonumber\\
        &&+\sum_{jk}T_{jk}X_j(VO)X_k(VO)\nonumber\\
        &&+\sum_{jkl}U_{jkl}X_j(VO)X_k(VO)X_l(VO),
\end{eqnarray}
where $x(BFT)$ is the $x$ coordinate at BFT and $X_{i}(VO)$ represents each vector element at $VO$.
The $x(BFT)$ was measured by BFT, while $x$, $y$, $x^{'}$, and $y^{'}$ in $X(VO)$ were determined from the local straight track reconstructed by BC3 and 4.
Thus, the beam momentum $\delta$ was obtained by solving Eq.~(\ref{eq:transferMT}) with respect to $\delta$.
If multi-track candidates remained, such events were recognized as multi-beam-particle events and rejected.
We lost 6\% of the total reconstructed events due to this rejection.

\subsubsection{SKS complex}
The local straight tracks were determined at the entrance and at the exit of the SKS magnet.
At this stage, accidental hits on the SFT were rejected by placing a timing gate of $\pm$3.5 ns on the SFT $x$ plane and a timing gate ranging from $-$8 ns to $+$5 ns for the SFT $u$ and $v$ planes.
In addition, the local straight tracks were selected based on the tracking $\chi^{2}$.
The momentum of a scattered particle was evaluated from a reconstructed trajectory inside the SKS magnet, by connecting local-track hit coordinates at the entrance and at the exit of the SKS magnet.
The SKS track was reconstructed using the fourth-order Runge-Kutta method \cite{RungeKutta}, with the magnetic field being calculated using \textsc{ansys}.
Multi-track events in which more than one track were found were rejected; overall, 0.2\% of the events were lost due to this rejection.

\subsection{Particle identification}
\label{sec:pid}
\subsubsection{Beam $\pi^{-}$ identification}
Electron contamination comprised the primary background in the $\pi^{-}$ beam.
GC was unable either to identify or to reject electrons in the $\pi^{-}$ beam correctly when its intensity is quite high, $12 \times 10^{6}$ pions per spill.
However, electrons did not comprise the background $K^{+}$ in the region of interest of the missing-mass spectra.
The number of beam particles was corrected by applying the electron contamination factor in the cross section analysis.

\subsubsection{$K^{+}$ identification}
\begin{figure}[htb]
  \includegraphics[width=8.5cm,clip]{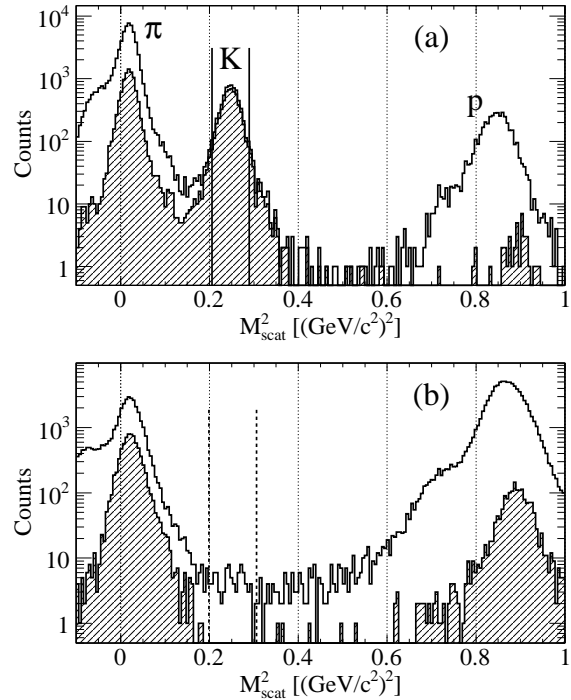}
  \caption{$M^{2}_{\mathrm{scat}}$ distributions of the scattered particles for the momentum ranges of (a) 750$\:<p_{\mathrm{SKS}}<\:$800 MeV/$c$ and (b) 900$\:<p_{\mathrm{SKS}}<\:$950 MeV/$c$. The non-hatched and hatched histograms indicate data obtained before and after the TOF $dE/dx$ gating, respectively. The solid lines around the $K^{+}$ peak in (a) represent the $M^{2}_{\mathrm{scat}}$ selection range, which is $2\sigma$ of the Gaussian function, at $p_{\mathrm{SKS}} = 775$ MeV/$c$. The dashed lines in (b) are the $M^{2}_{\mathrm{scat}}$ selection range at 925 MeV/$c$ extrapolated from the momentum region in which the $K^{+}$ peak is observed.}
  \label{fig:m2}
\end{figure}

Among the scattered particles, we must discriminate the $K^{+}$ from the $\pi^{+}$ and protons.
Note that high-momentum protons, which had a velocity larger than the $\beta$ threshold of LC, comprised the main background.
Mass-square information $M^{2}_{\mathrm{scat}}$ was used to identify $K^{+}$.
$M^{2}_{\mathrm{scat}}$ was calculated from the expression
\begin{equation}
  M^{2}_{\mathrm{scat}} = \left( \frac{p_{\mathrm{SKS}}}{\beta} \right)^{2} (1-\beta^{2}).
\end{equation}
Here, $p_{\mathrm{SKS}}$ is the momentum of the scattered particle, and $\beta$ was obtained from the time-of-flight and flight path length between BH2 and the TOF wall.
The $M^{2}_{\mathrm{scat}}$ distributions in the momentum ranges of 750$\:<p_{\mathrm{SKS}}<\:$800 MeV/$c$ and 900$\:<p_{\mathrm{SKS}}<\:$950 MeV/$c$ are shown in Fig.~\ref{fig:m2} (a) and (b), respectively, with non-hatched histograms.
The 900$\:<p_{\mathrm{SKS}}<\:$950 MeV/$c$ momentum range was the region in which the $\Lambda$ binding assumes unphysical high value.
It is apparent that the peak structures of the $\pi^{+}$ and protons do not exhibit simple Gaussian shapes and that the accompanying long tails extend below the $K^{+}$ mass region.
This suggests that additional methods to identify $K^{+}$ are necessary.

\begin{figure*}[htb]
  \includegraphics[width=14cm,clip]{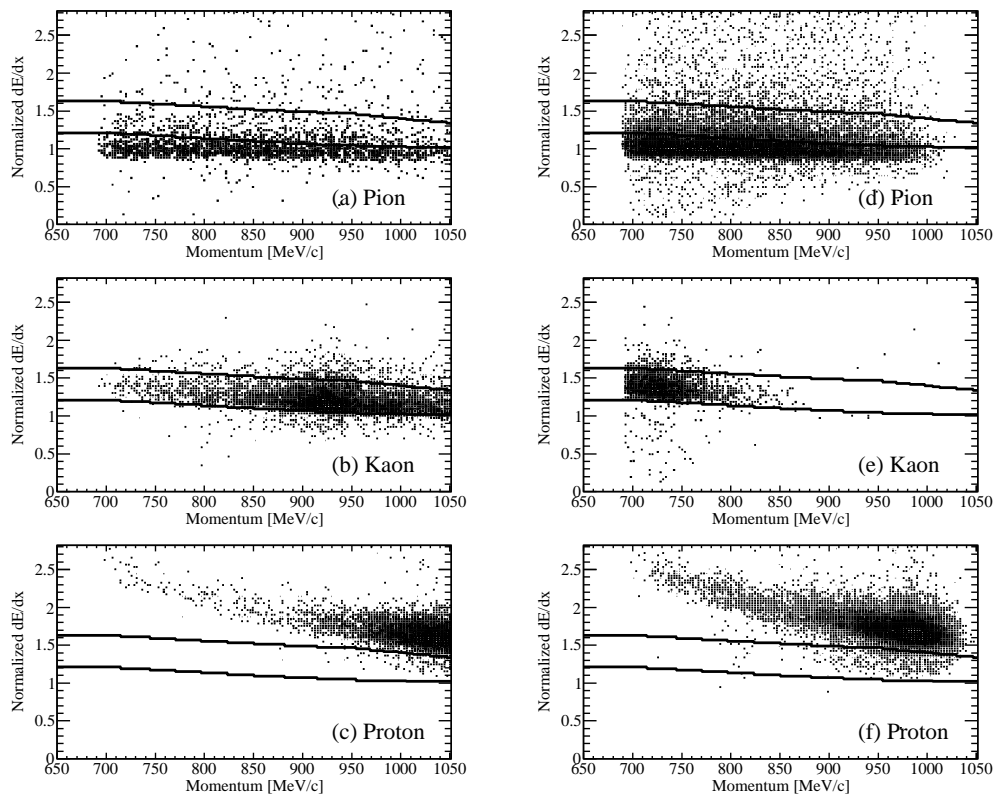}
  \caption{(a--c) Correlation plots between $dE/dx$ in the TOF wall segments and momenta of the scattered particles for particles selected using $M^{2}_{\mathrm{scat}}$ for data set D. (d--f) The analogous plots for data set A. The $dE/dx$ gating region for $K^{+}$ is indicated by the black lines. The gating bandwidth is constant below $960$ MeV/$c$, but narrows when the momentum is increased above $960$ MeV/$c$.}
  \label{fig:de}
\end{figure*}

We introduced a momentum-dependent $dE/dx$ gating for the signals from the TOF wall segments.
The correlations between $dE/dx$ and the momentum are plotted in Fig.~\ref{fig:de}. 
The vertical and horizontal axes indicate the $dE/dx$ in the TOF wall segment and the momentum of the scattered particles, respectively.
The peak position of $dE/dx$ for the $\pi^{+}$ was normalized to 1.
Figure~\ref{fig:de} (a), (b), and (c) shows the $dE/dx$ distributions for the $\pi^{+}$, $K^{+}$, and protons obtained from data set D, respectively.
The particles were identified using the $M^{2}_{\mathrm{scat}}$ information in order to create these plots.
Figure~\ref{fig:de} (d), (e), and (f) are the analogous plots created using data set A.
The solid lines represent the $dE/dx$ gating region for $K^{+}$.

We set the gating region according to the following procedures.
Specifically, the $dE/dx$ gating region was determined from data set D, the $\mathrm{CH_{2}}$ target data, because $K^{+}$ events were clearly observed in the high-momentum region for that data set, as shown in Fig.~\ref{fig:de} (b).
The $K^{+}$ momentum when a $\Sigma^{+}$ was produced at a beam momentum of 1.39 GeV/$c$ was roughly 900 MeV/$c$; this value is close to the $K^{+}$ momentum  when a $^{6}_{\Lambda}$H is produced with a 1.2-GeV/$c$ beam.
The $dE/dx$ distributions for the $\pi^{+}$, $K^{+}$, and protons were well separated up to 960 MeV/c in the analysis presented in this paper.
Therefore, the $K^{+}$ gating region was determined at 960 MeV/$c$ and the same gating width was applied to the momentum region below 960 MeV/$c$.
The averaged gating efficiency for $K^{+}$ below 960 MeV/$c$ was 85.3$\:\pm\:$1.3\%.
On the other hand, in the momentum region higher than 960 MeV/$c$, the $\pi^{+}$ and proton distributions exhibited a greater overlap with the $K^{+}$ distribution.
In such a high-momentum region, $K^{+}$ events via the $^{6}$Li($\pi^{-}, K^{+}$)X reaction are not expected, and then this region was used for the background level estimation.
In the present analysis, the $K^{+}$ distribution overlapping the proton one was removed from the gating region.
Thus, the gating region become narrower as the momentum increased.

The same gating region was applied to data set A, the $^{6}$Li data, as shown in Fig.~\ref{fig:de} (d), (e), and (f).
With the $dE/dx$ gating, the $\pi^{+}$ and proton contamination of the $K^{+}$ region was suppressed, as shown by the hatched histograms in Fig.~\ref{fig:m2} (a) and (b). 
Finally, we selected $K^{+}$ events in the $M^{2}_{\mathrm{scat}}$ distribution after applying the $dE/dx$ gating. 
In the momentum region in which the $K^{+}$ peak was apparent, the selection range was 2$\sigma$ of the Gaussian function for the $K^{+}$ peak.
The $M^{2}_{\mathrm{scat}}$ cut efficiency was estimated in this region and was 92.3$\:\pm\:$1.7\%.
The main contribution to the error was the uncertainty of the $\pi^{+}$ and proton contamination.
On the other hand, as no $K^{+}$ peak appeared in the high-momentum region, as shown in Fig.~\ref{fig:m2} (b), the selection range was extrapolated from the momentum region in which the $K^{+}$ peak is observed.

\subsection{Vertex reconstruction}
\begin{figure}[htb]
  \includegraphics[width=8.5cm,clip]{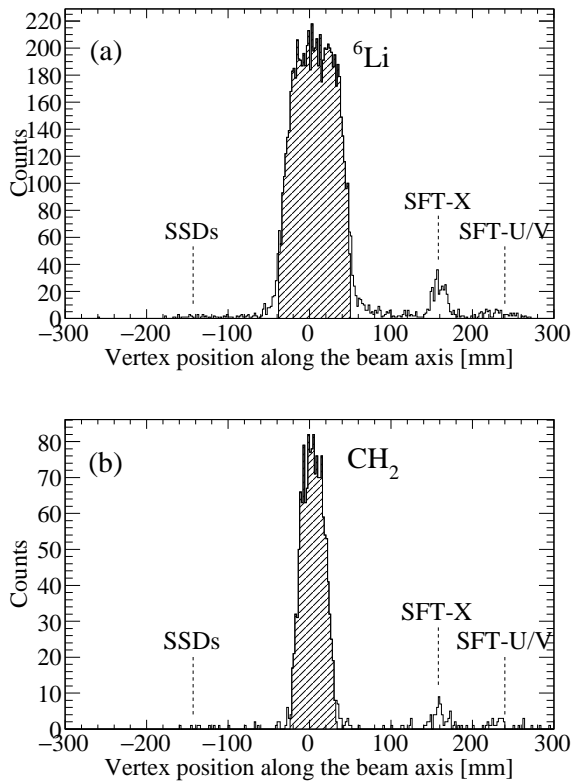}
  \caption{Reaction vertex distributions along the beam axis for data sets (a) A ($^{6}$Li target) and (b) D ($\mathrm{CH_{2}}$ target). The detector positions around the target are indicated by the dashed lines. The hatched region was selected as the target cut region.} 
  \label{fig:vtz}
\end{figure}

The reaction vertex was defined as the midpoint of the normal vector between the beam track and SKS track.
The local straight track reconstructed by BC3 and 4 was corrected by the SSD information and redefined as the beam track, because the experimental target was far from BC3 and 4, at roughly 1.2-m distance.
As the vertex resolution was affected by the multiple scattering in crossing BH2, the local straight track was extrapolated to the BH2 position, and the beam track was a straight line connecting the extrapolated track position and the hits on the SSDs.
For data set A, the distribution of the obtained vertex position along the beam axis for events in the angular region of  $4^{\circ}<\theta_{\pi K}<6^{\circ}$, where $\theta_{\pi K}$ is the reaction angle, are shown in Fig.~\ref{fig:vtz} (a).
The beam arrives from the left hand side in this figure and the center of the horizontal axis is the ideal target center.
Each detector position is represented by a dashed line in Fig.~\ref{fig:vtz}.
The $^{6}$Li target profile is clearly apparent at approximately the center of the figure.
However, the vertex distribution is slightly shifted from the center of the horizontal axis, because the target was actually installed somewhat downstream of the ideal position.
We selected an actual target length $+5$ mm around the central point of the vertex distribution, as indicated by the hatched region in Fig.~\ref{fig:vtz}.
The vertex selection efficiency for Fig.~\ref{fig:vtz} (a) is 96.1$\:\pm\:$0.6\%.

Figure~\ref{fig:vtz} (b) shows the reaction vertex distribution obtained from data set D, the polyethylene target data for the $\Sigma^{+}$ production.
In this case, we selected an actual target length $+10$ mm. 
In this case, the vertex selection efficiency is 98.1$\:\pm\:$0.8\%.

\subsection{Momentum calibration}
The momenta reconstructed by the two spectrometers were corrected by considering the energy loss and the correction functions of the two spectrometers, which were determined using the following procedures.
The energy loss in the $^{6}$Li target was estimated using a Monte Carlo simulation.
The validity of the simulation was confirmed by considering the measured energy loss in the target, which was obtained from data sets E and E', the beam through runs without and with the $^{6}$Li target, respectively.
The energy losses in the other materials such as detectors and other targets, were also estimated using the Monte Carlo simulation.

\begin{figure}[htb]
  \includegraphics[width=8.5cm,clip]{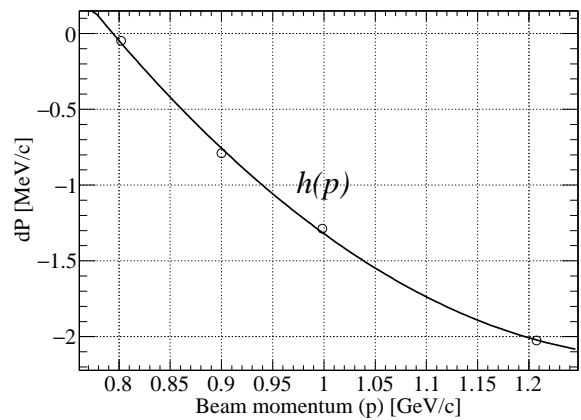}
  \caption{Mean value of $dP$ as a function of the beam momentum. The data points were obtained from data set E, the $\pi^{+}$ beam through run. The vertical and horizontal axes indicate the mean value of $dP$ and the beam momentum, respectively. The solid line is the quadratic fitting function, $h(p)$.}
  \label{fig:bt}
\end{figure}

The momentum correction functions for the beam line spectrometer and the SKS complex, $f_{\mathrm{beam}}(p)$ and $g_{\mathrm{SKS}}(p)$, were respectively defined as
\begin{eqnarray}
  f_{\mathrm{beam}}(p) &=& \alpha p + \beta, \\
  g_{\mathrm{SKS}}(p)  &=& Ap^{2} + Bp + C.
\end{eqnarray}
Here, we assumed that the SKS complex had the 2nd order non-linearity while the beam line spectrometer did not.
The parameters of these correction functions were obtained via the following procedures.
From data set E, the $\pi^{+}$ beam through runs, the $\pi^{+}$ beam momentum was measured by the two spectrometers.
The beam momentum difference, $dP = p'_{\mathrm{SKS}}-p'_{\mathrm{beam}}$, was defined, where $p'_{\mathrm{SKS}}$ and $p'_{\mathrm{beam}}$ are the energy-loss corrected momenta obtained by the SKS complex and the beam line spectrometer, respectively.
Figure~\ref{fig:bt} shows the mean values of $dP$ as a function of the beam momentum.
The correlation was fitted using a quadratic function with residual values smaller than 40 keV/$c$.
This fitting function contained contributions from the two spectrometers.
Then, the relation between the fitting function in Fig.~\ref{fig:bt} and the correction functions for the spectrometers was expressed as
\begin{eqnarray}
  h(p) &=& g_{\mathrm{SKS}}(p) - f_{\mathrm{beam}}(p) \nonumber \\
       &=& Ap^{2} + (B-\alpha)p + (C-\beta),
\end{eqnarray}
where $h(p)$ is the fitting function shown in Fig.~\ref{fig:bt}.
At this stage, we were unable to separate the correction functions for the beam line spectrometer and the SKS complex.
The fitting parameters were $A$, ($B-\alpha$), and ($C-\beta$).
In the following stage, the absolute momenta were obtained by adjusting $\alpha$ and $\beta$ using the $\Sigma^{\pm}$ missing masses extracted from data sets C and D.

An additional constant value $\gamma$ was introduced when the $\pi^{-}$ beam was used, because the beam polarity change produces an offset on the momentum value.
Thus, we were required to determine three parameters, $\alpha$, $\beta$, and $\gamma$, to reproduce the $\Sigma^{\pm}$ masses.
However, several combinations of these parameters were permitted, because the number of calibration points was less than the number of parameters.
Therefore, we considered two extreme cases, with $\alpha = 0$ or $\beta = 0$, in order to estimate the systematic error of the missing mass originating from the $\alpha$ and $\beta$ determination around the bound region of $^{6}_{\Lambda}$H.
Here, we calculated the missing mass of the $^{6}$Li($\pi^{-}, K^{+}$)X reaction for the missing-mass error estimation by assuming that the ground state of $^{6}_{\Lambda}$H was produced.
The beam and SKS momenta, which we assumed, were corrected using two different correction functions for $\alpha = 0$ and $\beta = 0$.
As a result, the mass difference between these two cases was 310 keV/$c^{2}$.
In this analysis, the error of the absolute scale of the missing mass was dominated by this momentum correction uncertainty.
Furthermore, other mass scale errors, namely, the energy-loss uncertainty in the target materials and the residual between the data and $h(p)$ were taken into account.
The mass scale error from the energy-loss uncertainty was estimated as follows.
We re-estimated the parameters of the correction functions for the two spectrometers by changing the energy loss in the target by $\pm 5$\%.
The mass scale error was obtained as the missing-mass difference between the original analysis result and the result obtained with the re-estimated correction functions, and the error was 90 keV/$c^{2}$.
The maximum value of the residual between $dP$ and $h(p)$ was 32 keV/$c$.
This corresponded to 26 keV/$c^{2}$ in the missing mass assuming that the kaon momentum was incorrect.
Finally, the overall missing-mass scale error in this experiment was 350 keV/$c^{2}$ in the vicinity of the bound region of $^{6}_{\Lambda}$H.

\subsection{Raw missing-mass spectrum}
The missing mass of the $^{6}$Li($\pi^{-}, K^{+}$)X reaction was defined as
\begin{equation}
  M_{X}\mathalpha{=}\sqrt{(E_{\pi}\mathalpha{+}M_{\mathrm{tgt}}\mathalpha{-}E_{K})^{2}\mathalpha{-}({p_{\pi}}^{2}\mathalpha{+}{p_{K}}^{2}\mathalpha{-}2p_{\pi}p_{K}cos(\theta_{\pi K}))}.  
\end{equation}
Here, $p_{\pi}$ and $p_{K}$ are the corrected momenta of $\pi^{-}$ and $K^{+}$ at the reaction point, respectively, $E_{\pi}$ and $E_{K}$ are the total energies of $\pi^{-}$ and $K^{+}$ calculated from the corrected three momenta, respectively, $M_{\mathrm{tgt}}$ is the mass of the target, and $\theta_{\pi K}$ is the reaction angle.

\begin{figure}[htb]
  \includegraphics[width=8.5cm,clip]{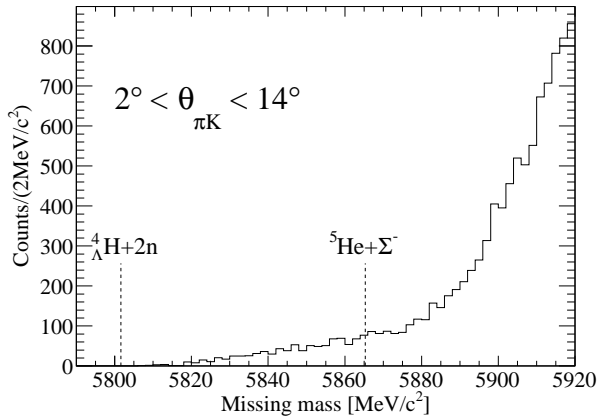}
  \caption{Missing-mass spectrum of the $^{6}$Li($\pi^{-}, K^{+}$)X reaction in the angular range of $2^{\circ}<\theta_{\pi K}<14^{\circ}$. Two mass thresholds, $^{4}_{\Lambda}{\mathrm{H}} + 2n$ and  $^{5}{\mathrm{He}} + \Sigma^{-}$, are indicated by the dashed lines. The horizontal axis bin width is 2 MeV/$c^{2}$.}
  \label{fig:h6l_mm_count}
\end{figure}

We calculated the missing mass of the $^{6}$Li($\pi^{-}, K^{+}$)X reaction in the angular range of $2^{\circ}<\theta_{\pi K}<14^{\circ}$ and obtained a spectrum up to 5920 MeV/$c^{2}$, as shown in Fig.~\ref{fig:h6l_mm_count}.
The two dashed lines in the figure represent the mass thresholds of $^{4}_{\Lambda}{\mathrm{H}} + 2n$ and $^{5}{\mathrm{He}} + \Sigma^{-}$.
The horizontal axis indicate the missing mass in units of 2 MeV/$c^{2}$.
Owing to the improvement in the tracking efficiencies, the number of events retained in the present analysis was 20\% larger than that of the previous one.

\subsection{Missing-mass resolution}
The experimental missing-mass resolution was estimated using the missing-mass spectrum of the $^{12}_{\Lambda}$C production obtained from data set B.
By taking into account the target thickness of $^{6}$Li, the missing-mass resolution was primarily determined by considering the energy-loss straggling in the target.
Therefore, a graphite target of 3.6 g/$\mathrm{cm^{2}}$ was used to estimate the missing-mass resolution, because the energy-loss straggling in the graphite target was expected to be almost identical to that in the $^{6}$Li target.

\begin{figure}[htb]
  \includegraphics[width=8.5cm,clip]{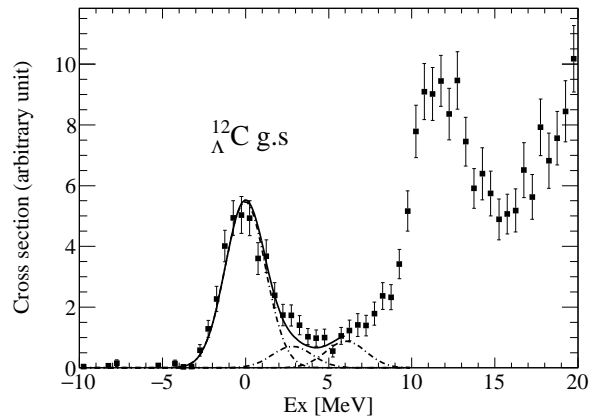}
  \caption{Excitation energy spectrum of the $^{12}_{\Lambda}$C hypernucleus. The peak position of the first Gaussian function was set to 0 MeV. The low-lying states were fitted with three Gaussian functions. The peak separation and their relative amplitude were fixed. The dot-dashed lines represent each Gaussian function and the solid line is their sum. The missing-mass resolution was found to be 2.9 MeV/$c^{2}$ (FWHM).}
  \label{fig:c12l}
\end{figure}

Figure~\ref{fig:c12l} shows the excitation energy spectrum of the $^{12}_{\Lambda}$C hypernucleus.
Here, the low-lying states were fitted with three Gaussian functions having the same width.
The solid line is the fitting result and the dot-dashed line shows each Gaussian function.
The relative peak position and the relative amplitude of Gaussian functions were fixed by existing $\gamma$-ray spectroscopy \cite{Hosomi} and missing-mass spectroscopy \cite{hyp4} results, respectively.
The peak position of the first Gaussian function was set to 0 MeV by the fit.
A missing-mass resolution of $2.9$ MeV/$c^{2}$ full width at half maximum (FWHM) was obtained.

\subsection{Cross section}
The double-differential cross section, $\frac{d^{2}\sigma}{d\Omega dM}$, was defined as
\begin{eqnarray}
  \frac{d^{2}\sigma}{d\Omega dM} =&& \frac{1}{N_{\mathrm{target}}}\cdot \frac{1}{N_{\pi} \cdot \epsilon_{\mathrm{K1.8}}} \cdot \frac{N_{K}}{\epsilon_{\mathrm{SKS}}} \cdot \frac{1}{\epsilon_{\mathrm{vertex}}}\nonumber\\
  && \times \frac{1}{\epsilon_{\mathrm{DAQ}}} \cdot \frac{1}{f_{K}} \cdot \frac{1}{d\Omega} \cdot \frac{1}{dM},
\end{eqnarray}
\begin{eqnarray*}
  N_{\pi}          &=& N_{\mathrm{beam}} \cdot f_{\pi}, \\
  N_{\mathrm{target}} &=& f_{\mathrm{enrich}} \cdot \frac{(\rho x)\cdot N_{A}}{A}.
\end{eqnarray*}
Here, $N_{\mathrm{target}}$, $N_{\mathrm{beam}}$, and $N_{K}$ represent the numbers of $^{6}$Li nuclei in the target, the beam particles triggered by BH2, and the $K^{+}$ identified in the raw missing-mass analysis, respectively.
$N_{\mathrm{target}}$ was determined based on the fraction of $^{6}$Li nuclei in the Li target ($f_{\mathrm{enrich}}$), the target thickness ($\rho x$), Avogadro's number ($N_{A}$), and the atomic mass ($A$).
$d\Omega$ and $dM$ denote the effective solid angle and the missing-mass bin size in the histogram, respectively.
The other $\epsilon$'s and $f$'s are the efficiencies and correction factors, respectively.
The $K^{+}$ survival ratio, $f_{K}$, contains the $K^{+}$ decay and absorption factors. 
These efficiencies and factors in the angular range of $4^{\circ} < \theta_{\pi K} < 6^{\circ}$ are listed in Table \ref{tb:eff_table} as an example.

We describe the details of the efficiencies and the correction factors in the following.

\begin{table}[htb]
  \caption{Efficiencies and correction factors in the angular range of $4^{\circ} < \theta_{\pi K} < 6^{\circ}$ used in the cross section calculation.}
  \begin{tabular*}{\columnwidth}{@{\extracolsep{\fill}}cccc} \hline \hline
                                  & $^{6}$Li (data set A)      & $\Sigma^{-}$ (data set C)      & $\Sigma^{+}$ (data set D) \\ 
                                  & (\%)                       & (\%)                           & (\%) \\ \hline
    $f_{\pi}$                     & 80.8$\:\pm\:$0.6           & 82.9$\:\pm\:$0.5               & 83.1$\:\pm\:$0.6 \\ 
    $\epsilon_{\mathrm{K18}}$     & 50.8$\:\pm\:$0.9           & 84.5$\:\pm\:$0.8               & 70.3$\:\pm\:$0.8 \\ 
    $\epsilon_{\mathrm{SKS}}$     & 50.5$\:\pm\:$2.0           & 48.3$\:\pm\:$1.8               & 56.5$\:\pm\:$1.9 \\ 
    $\epsilon_{\mathrm{vertex}}$  & 94.4$\:\pm\:$0.6           & 97.1$\:\pm\:$1.0               & 97.5$\:\pm\:$0.8 \\ 
    $\epsilon_{\mathrm{DAQ}}$     & 75.7$\:\pm\:$1.2           & 74.7$\:\pm\:$0.8               & 64.4$\:\pm\:$0.8 \\ 
    $f_{K}$                       & 40.9$\:\pm\:$0.8           & 46.7$\:\pm\:$0.8               & 46.7$\:\pm\:$0.8 \\ \hline \hline
  \end{tabular*}
  \label{tb:eff_table}
\end{table}

\subsubsection{Beam normalization}
The main background in the $\pi^{-}$ beam due to the electron contamination.
As discussed in Sec. \ref{sec:pid}, we were unable to identify electrons in the event-by-event analysis.
Moreover, as $N_{\mathrm{beam}}$ corresponds to the number of charged particles triggered by BH2, this value was corrected using the electron contamination and the $\mu^{-}$ contamination factors in order to estimate the number of $\pi^{-}$, $N_{\pi}$.
The electron contamination fraction was estimated using data obtained under the low-intensity beam condition, and was 11.3$\:\pm\:$0.2\%.
The other contribution to the background were $\mu^{-}$ coming from $\pi^{-}$ decays in the analyzer magnet of the beam line spectrometer.
The $\mu^{-}$ contamination at the target position was estimated using \textsc{decay-turtle} \cite{TURTLE} and was 2.2\%.
In addition, only 93.1$\:\pm\:$0.6\% of the beam impinged on the target, because of its finite horizontal size.
On the other hand, the $^{6}$Li target was fully irradiated by the vertical distribution of the beam profile.
Thus, the overall beam normalization factor ($f_{\pi}$) was 80.8$\:\pm\:$0.6\%.

\subsubsection{Beam analysis efficiency}
The beam analysis efficiency was a product of the momentum reconstruction efficiency and of the SSD analysis efficiency for the vertex reconstruction.
The beam momentum reconstruction process included timing selection of BH2 and BFT, $\chi^{2}$ selection of the BC3 and 4 local tracking, and multi-beam-particle rejection.
At this stage, 82.0$\:\pm\:$1.0\% of the events were retained and recognized as single-beam-particle events.

SSD was the key detector in the beam line to improve the vertex resolution, as it suppressed the background from materials other than the target.
For this analysis, the number of hit on SSD must be only one.
However, multi-hits remained, because of the high-rate condition even after the background rejection analysis.
Then, the analysis efficiency of SSD was 61.2$\:\pm\:$0.8\% for data set A, because of following rejection of the SSD multi-hit events, where the reaction vertex position was not uniquely determined.
Thus, the beam analysis efficiency $\epsilon_{\mathrm{K18}}$ was 50.8$\:\pm\:$0.9\%.

\subsubsection{SKS analysis efficiency}
The SKS analysis efficiency was determined by the local tracking efficiencies, the momentum reconstruction efficiency, the trigger-counter detection efficiencies, and the $K^{+}$ identification efficiency described in Sec.~\ref{sec:pid}.

\begin{figure}[htb]
  \includegraphics[width=8.5cm,clip]{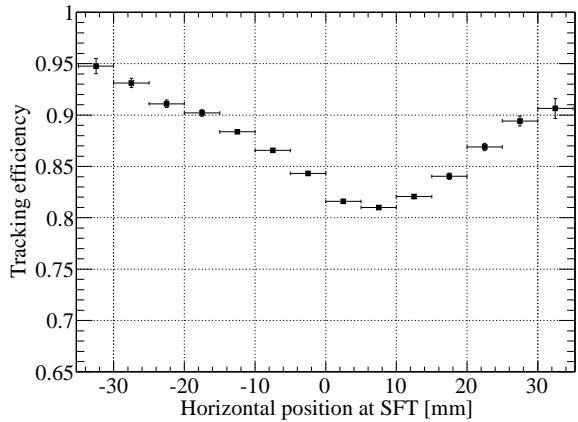}
  \caption{Horizontal position dependence of SFT-SDC2 local tracking efficiency. The origin of the horizontal axis corresponds to the detector center. Decreased efficiency is observed near the center, because of the beam focus.}
  \label{fig:eff_sdcin}
\end{figure}

The local tracking efficiency of SFT-SDC2 had a horizontal position dependence. 
As these detectors were placed immediately behind the target on which the beam was focused, the tracking efficiency in the vicinity of the detector center decreased under high-intensity beam condition.
Figure~\ref{fig:eff_sdcin} shows the spatial position dependence of the local SFT-SDC2 tracking efficiency.
The abscissa indicates the horizontal position of the beam trajectory at the SFT $x$ plane and the origin is the SFT center.
A decrease in efficiency is apparent near the SFT center.
Thus, an efficiency table depending on the horizontal position was used in the cross section calculation.
A systematic error of 2\% was determined from the efficiency time variation.
On the other hand, the local tracking efficiency of SDC3 and 4 was 97.9$\:\pm\:$1.2\% with no position dependence. 
The momentum reconstruction efficiency was obtained as a function of the scattering angle.
A maximum efficiency variation of 2\% in the angular range of $2^{\circ} < \theta_{\pi K} < 16^{\circ}$.
Its typical value was 95.4$\:\pm\:$0.3\%.

TOF, LC, and LAC were the trigger counters in the SKS complex.
The detection efficiency of LC was obtained as 97.8$\:\pm\:$0.1\% for $K^{+}$, while the TOF detection efficiency was assumed to be 100\%.
In addition, 13.7$\:\pm\:$1.4\% of the $K^{+}$ were erroneously suppressed by LAC at the trigger level.

As described in Sec.~\ref{sec:pid}, $K^{+}$ events were identified using two analysis cuts, the TOF $dE/dx$ gating and the $M^{2}_{\mathrm{scat}}$ selection.
The efficiency of the $dE/dx$ gating obtained from data set A was 85.3$\:\pm\:$1.3\%.
The differences between the efficiencies estimated from data sets A, C, and D were regarded as a systematic error.
The $K^{+}$ selection efficiency using  $M^{2}_{\mathrm{scat}}$ was estimated by counting the remaining number of $K^{+}$ and was 92.3$\:\pm\:$1.7\%.
In this estimation, the main contribution to the error was uncertainty as to whether the side band events around the $K^{+}$ peak in Fig.~\ref{fig:m2} were genuinely corresponded to $K^{+}$.

Finally, $\epsilon_{\mathrm{SKS}}$ was calculated as by multiplying all efficiencies and was 50.5$\:\pm\:$2.0\%.

\subsubsection{Vertex efficiency}
\begin{figure}[htb]
  \includegraphics[width=8.5cm,clip]{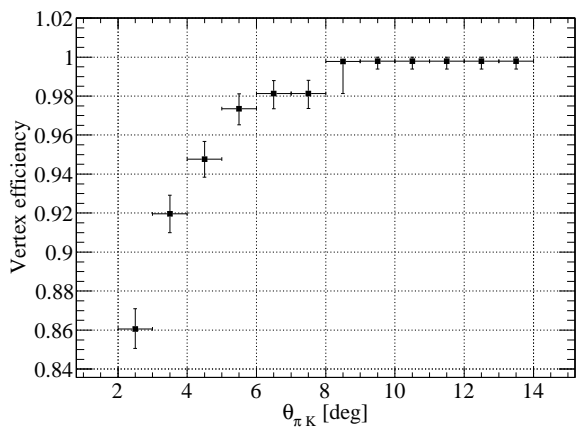}
  \caption{Vertex selection efficiency as a function of the reaction angle, $\theta_{\pi K}$.}
  \label{fig:eff_vertex}
\end{figure}

The vertex selection efficiencies were estimated using template fitting to the vertex distributions in every one-degree reaction angle step.
The efficiency was obtained as a function of the reaction angle by integrating the template function, which was a uniform distribution convoluted with a Gaussian function.
Figure~\ref{fig:eff_vertex} shows the vertex selection efficiency for the $^{6}$Li target.
As the vertex distribution was broader at the forward angle, the efficiency was smaller.
In addition, events featuring a closest distance between the beam and the kaon trajectories less than 3 mm were selected as good vertex events.
This selection efficiency was 98.2$\:\pm\:$0.1\%.
The vertex correction factor $\epsilon_{\mathrm{vertex}}$ was calculated as a product of the vertex selection and closest distance selection efficiencies.

\subsubsection{DAQ efficiency}
A typical value of the data acquisition (DAQ) efficiency during the physics run was 78\%.
In addition, some of the $K^{+}$ events were suppressed by the 2nd-level trigger as a result of misidentification, and its correction factor was 97.2$\:\pm\:$0.1\%.
Then, the DAQ correction factor $\epsilon_{\mathrm{DAQ}}$ was estimated to be 75.7$\:\pm\:$1.2\%.

\subsubsection{$K^{+}$ survival ratio}
$K^{+}$ were lost as a result of their absorption between the target and the LC, and because of their decay before traversal of the LAC.
The $K^{+}$ absorption in the materials between the target and the LC was studied using a simulation based on the G\textsc{eant}4 package \cite{G4} and was 3.1\%.
Thus, the corresponding correction factor was 96.9$\:\pm\:$0.5\%.
On the other hand, event-by-event correction was applied for the $K^{+}$ decay by considering $\beta$ and the flight path length.
A typical values of 40\% was obtained for data set A.
These results are listed as $f_{K}$ in Table \ref{tb:eff_table}.

\subsubsection{Acceptance correction}
The effective solid angle of the SKS complex ($d\Omega$) was estimated using a simulation based on the G\textsc{eant}4 package.
The multiple scattering effect and the energy loss were taken into account, but the decay process was not activated during the simulation.
$K^{+}$ were generated uniformly in the center-of-mass (C.M.) system.
A realistic beam profile was reproduced in order to determine the effective solid angle as a function of the reaction angle, $\theta_{\pi K}$.
This solid angle in each $dM$ bin was defined as
\begin{equation}
\label{eq:solid_angle}
  d\Omega(dM) = 2\pi (cos\theta_{1} - cos\theta_{2}) \left< \frac{N_{\mathrm{a}}}{N_{\mathrm{g}}} \right>.
\end{equation}
Here, $\theta_1$ and $\theta_2$ are the end points of the angular range.
Further, $N_{\mathrm{a}}$ and $N_{\mathrm{g}}$ are the number of $K^{+}$ accepted and generated in the simulation, respectively.
The table comprised of the $N_{\mathrm{a}}/N_{\mathrm{g}}$ ratios is defined as the SKS-complex acceptance table.
The $N_{\mathrm{a}}/N_{\mathrm{g}}$ ratio was averaged over the scattered particle momentum and the reaction angle ($\theta$) in a given $dM$ bin, as indicated in Eq.~(\ref{eq:solid_angle}).
The cross section was calculated using this effective solid angle in an event-by-event manner, by looking at the acceptance table according to the reconstructed $K^{+}$ momentum and $\theta_{\pi K}$.

There were two different sources of systematic error in the acceptance correction procedure.
One was the finite mesh size effect.
This uncertainty depended on the table mesh size and was sizeable in the region in which the $p$ and $\theta$ dependencies were large.
The other error source was the acceptance edge effect, which was found in the low-momentum region in the angular range of $4^{\circ} < \theta_{\pi K} < 8^{\circ}$.
The SKS acceptance for the events in which the outgoing particles were scattered to the left hand side, with respect to the beam direction, went to 0 in this angular region.
On the contrary, it maintained its value when particles are scattered to the right hand side.
The events lying in the missing-mass range of 5890--5920 MeV/$c^{2}$ and the angular range of $4^{\circ} < \theta_{\pi K} < 8^{\circ}$ were affected by the acceptance edge effect on the left side of the SKS complex.
Thus, larger systematic errors were set in this kinematical region with respect to the other ones.

\subsection{Systematic errors}
The errors listed in Table \ref{tb:eff_table} are the systematic errors of each correction factor and efficiency.
The total systematic error of the present analysis was obtained by summing all systematic and acceptance errors described in the previous subsection.
Hence, the systematic errors in the angular range of $2^{\circ} < \theta_{\pi K} < 14^{\circ}$ surrounding the bound region were typically 5.1\%.
This value gradually increased with increasing missing mass (from the initial value of 5890 MeV/$c^{2}$) and reached 7\% at 5896 MeV/$c^{2}$, because of the edge effect of the acceptance correction.

\section{Elementary cross section}
\label{sec:elem}
One of the theoretical approaches to the analysis of the $\Lambda$-hypernuclei production via the DCX reaction consists in fitting the experimental spectrum with the DWIA calculation result.
In the DWIA calculation, the knowledge of cross section of the elementary process, $\pi^{\pm} p \rightarrow K^{+} \Sigma^{\pm}$, is required to calculate the $\Lambda$-hypernuclei production cross section.
Actually, the $\Lambda$ and $\Sigma$ particles may be strongly coupled as a result of the $\Lambda$-$\Sigma$ mixing in the $\Lambda$-hypernuclear state produced via the one-step process of the DCX reaction.
Then, a precise determination of the elementary cross section is necessary for reliable calculation of the $^{6}$Li($\pi^{-}, K^{+}$)X reaction.
We estimated the elementary cross sections of the $\pi^{\pm} p \rightarrow K^{+} \Sigma^{\pm}$ reactions from data sets C and D according to the described above procedures.

As data sets C and D are independent of data set A, we re-estimated the correction factors in the cross section calculations for data sets C and D (Table \ref{tb:eff_table}).
The DCX reaction was used for data sets A and C; however, the NCX reaction was used in the case of data set D.
The beam intensity in the measurement of the NCX reaction was one third of that in the case of the DCX reaction.
Owing to the lower beam intensity, the SSD analysis efficiency was higher in data set D.
On the other hand, SSD did not work well and it was not present in data set C, because of a hardware problem.
Thus, the $\epsilon_{\mathrm{K18}}$ of data set C did not contain the SSD efficiency.
Apart from the SSD efficiencies, the other correction factors and efficiencies were almost identical to those of data set A.
The systematic errors for each correction factor were also estimated and are summarized in Table \ref{tb:eff_table}.
These errors were estimated using the same approach as for data set A, except for the treatment of the SKS acceptance errors.
The $K^{+}$ momentum was higher and its region of interest was smaller for both data sets C and D compared to those of data set A.
Thus, the SKS acceptance edge caused a large systematic error in data set A, while this error was negligible in data sets C and D.
In addition, the finite mesh size effect of the acceptance table was also negligible, owing to the high statistics of the $\pi^{\pm} p \rightarrow K^{+} \Sigma^{\pm}$ events.
However, in data set C the production cross section of $\pi^{-} p \rightarrow K^{+} \Sigma^{-}$ for scattering to the left differed systematically from those of the scattering to the right as seen from the beam direction.
This difference was not observed for data sets A and D.
Thus, the systematic error of the SKS acceptance for data set C was conservatively estimated from the maximum difference between the left and right scattering, and was roughly 10\%.

\begin{figure*}[htb]
  \begin{center}
  \includegraphics[width=14cm,clip]{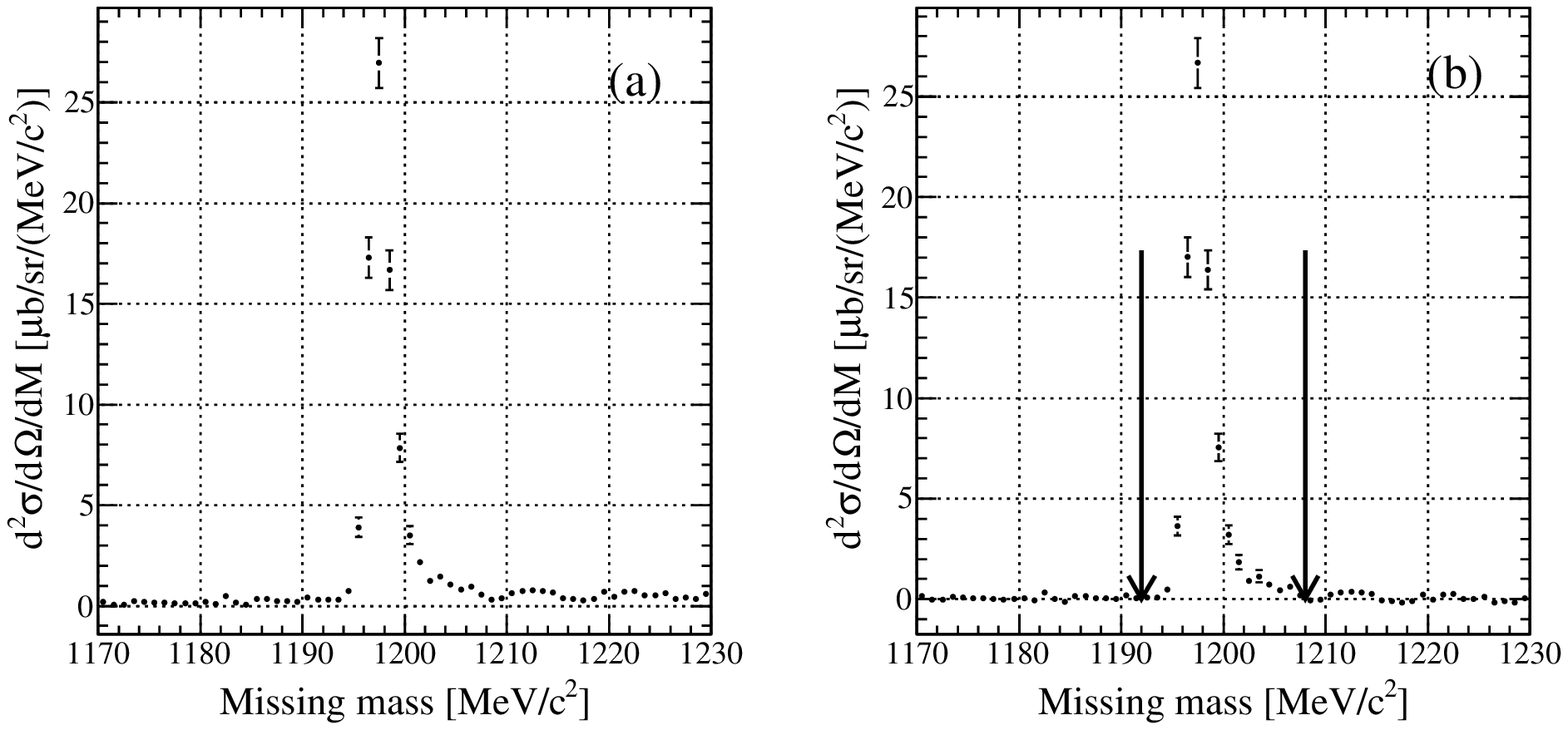}
  \caption{Double-differential cross sections of the $\pi^{-} p \rightarrow K^{+} \Sigma^{-}$ reaction at 1.39-GeV/$c$ beam momentum in $4^{\circ} < \theta_{\pi K} < 6^{\circ}$ angular range in (a) lab system and (b) after subtracting background. The arrows indicate the integration range used to estimate the cross section.}
  \label{fig:mm_sigmam}
  \end{center}
\end{figure*}

\begin{figure*}[htb]
  \includegraphics[width=14cm,clip]{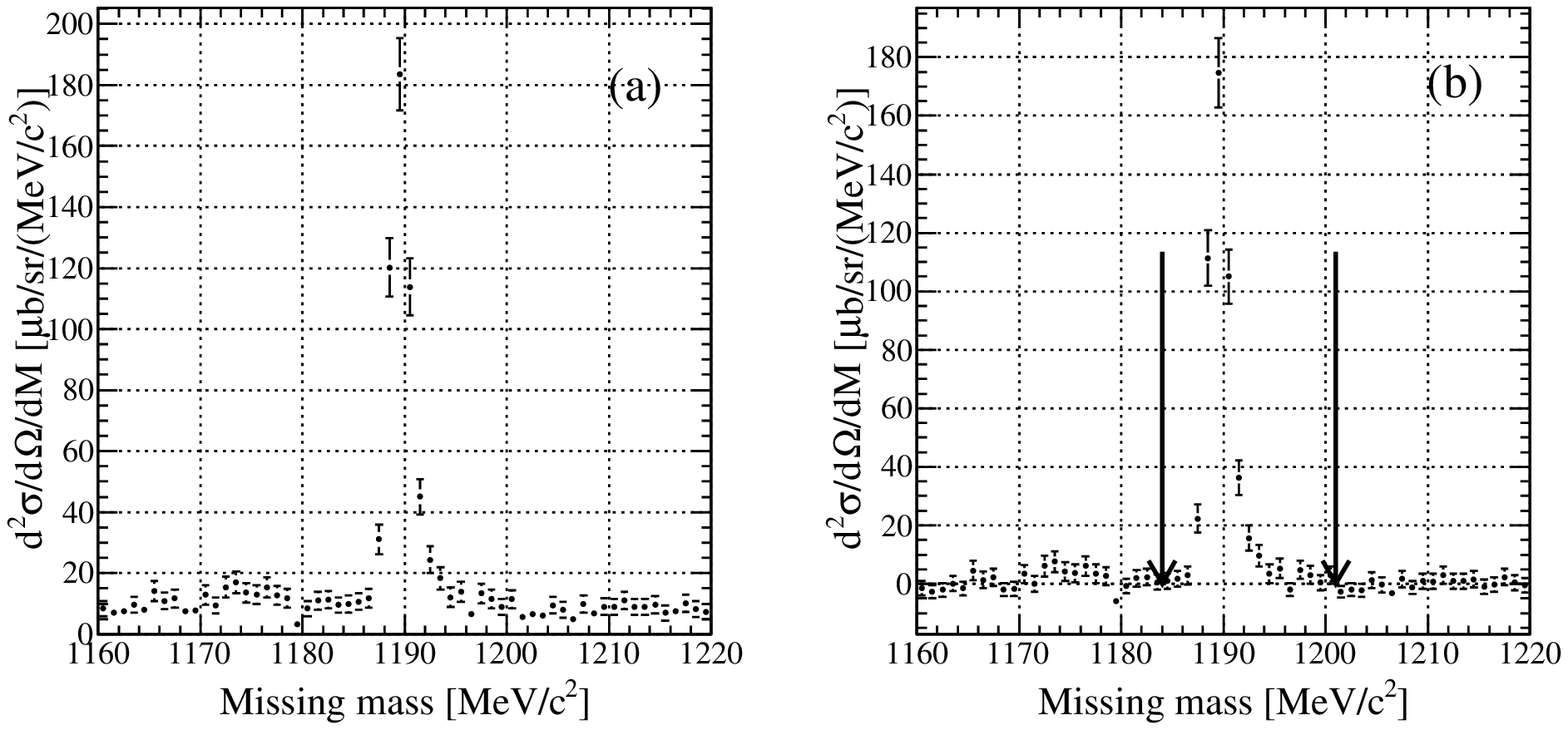}
  \caption{Double-differential cross sections of the $\pi^{+} p \rightarrow K^{+} \Sigma^{+}$ reaction at 1.39-GeV/$c$ beam momentum in $4^{\circ} < \theta_{\pi K} < 6^{\circ}$ angular range in (a) lab system and (b) after subtracting background. The arrows indicate the integration range used to estimate the cross section.}
  \label{fig:mm_sigmap}
\end{figure*}

\begin{figure}[htb]
  \includegraphics[width=8.5cm,clip]{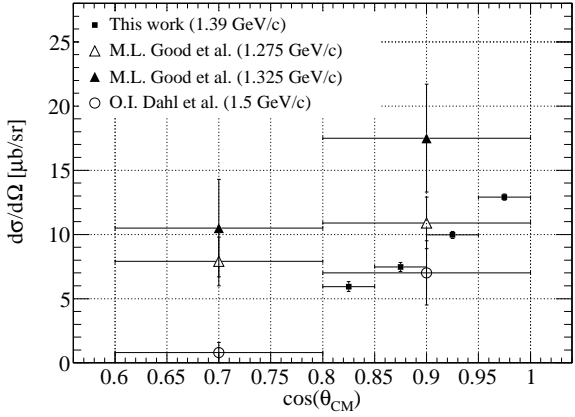}
  \caption{Differential cross section of the $\Sigma^{-}$ production in C.M. system and those from the past experiments. Our data are indicated by filled squares. The data points indicated by open and filled triangles are from Ref.~\cite{cs_sigmam_1} and the open circles correspond to data from Ref.~\cite{cs_sigmam_2}.}
  \label{fig:cs_sigmam_cm}
\end{figure}

\begin{figure}[htb]
  \includegraphics[width=8.5cm,clip]{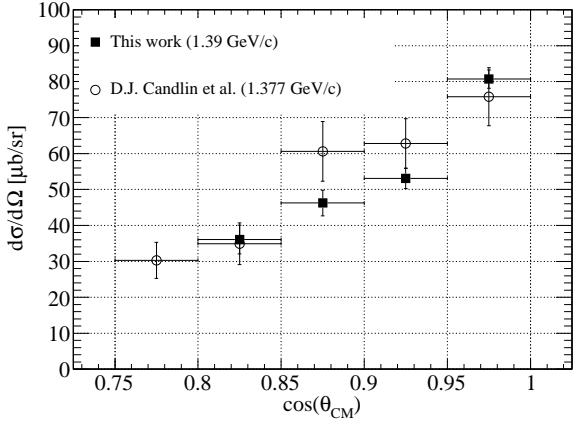}
  \caption{Differential cross section of the $\Sigma^{+}$ production in C.M. system and that from the past experiment \cite{cs_sigmap_1}. Our data and those from Ref.~\cite{cs_sigmap_1} are indicated by filled squares and open circles, respectively.}
  \label{fig:cs_sigmap_cm}
\end{figure}

The $\Sigma^{-}$ and $\Sigma^{+}$ double-differential cross section spectra in the angular range of $4^{\circ} < \theta_{\pi K} < 6^{\circ}$ in the lab system are shown in Fig.~\ref{fig:mm_sigmam} (a) and Fig.~\ref{fig:mm_sigmap} (a), respectively.
These cross section spectra also contain the contribution from the C($\pi^{-}, K^{+}$)X reaction, because the $\mathrm{CH_{2}}$ target was used.
This spectrum was fitted with double Gaussian functions for the signal and a linear function in order to estimate the background distribution from the carbon.
The second Gaussian function was used to simulate the tail structure of the $\Sigma$ peak.
The spectra obtained following subtraction of the carbon contribution are shown in Fig.~\ref{fig:mm_sigmam} (b) and Fig.~\ref{fig:mm_sigmap} (b). 
The production cross sections of $\Sigma^{\pm}$ were obtained by integrating the double-differential cross section in the integration range indicated by arrows in Fig.~\ref{fig:mm_sigmam} (b) and Fig.~\ref{fig:mm_sigmap} (b).
The standard deviation of the number of $\Sigma$ events, $\sqrt{N_{\Sigma}}$, and the error of subtracted area, which originated from the fitting, were incorporated into the statistical errors, while the integration range dependence was included in the systematic error.
The cross sections of $\pi^{\pm} p \rightarrow K^{+} \Sigma^{\pm}$ at a beam momentum of 1.39 GeV/$c$ in the C.M. system are plotted in Figs.~\ref{fig:cs_sigmam_cm} and \ref{fig:cs_sigmap_cm}, together with past experimental data.
The numerical values of the cross sections of the $\pi^{\pm} p \rightarrow K^{+} \Sigma^{\pm}$ reactions in the lab and C.M. systems are listed in Appendix~\ref{AppA}.

In the case of the $\Sigma^{-}$ production (Fig.~\ref{fig:cs_sigmam_cm}), the present result is plotted together with past experimental data obtained at beam momenta of 1.275 and 1.325 GeV/$c$ \cite{cs_sigmam_1} and at a beam momentum of 1.5 GeV/$c$ \cite{cs_sigmam_2}.
The error bars show the statistical errors only.
It is clear that the error bars of the present result are smaller than those of the past experimental data.
The previous experimental data indicate that the cross section of the $\pi^{-}p \rightarrow K^{+} \Sigma^{-}$ reaction decreased when the beam momentum increased.
Examination of the data points at cos$\theta$ of approximately 0.9 indicates that the present result agrees with this trend.
We obtained the production cross section of the $\pi^{-}p \rightarrow K^{+} \Sigma^{-}$ reaction at 1.39 GeV/$c$ in the angular range of $0.8 < $ cos$\theta < 1.0$.

In the case of $\Sigma^{+}$ production (Fig.~\ref{fig:cs_sigmap_cm}), as the beam momentum of the present experiment is close to that of the past experiment, the angular distributions of the cross sections can be compared directly.
The statistical errors of the present data are slightly superior or identical to those of the past experimental data.
The angular distribution of the present result is consistent with the past data \cite{cs_sigmap_1}, within the error bars.

\section{Results and discussion}
In this section, we show the obtained cross-section spectrum of the $^{6}$Li($\pi^{-}, K^{+}$)X reaction and we discuss both the $\Lambda$ bound and continuum regions.

\subsection{$\Lambda$ bound region}
\begin{figure}[htb]
  \includegraphics[width=8.5cm,clip]{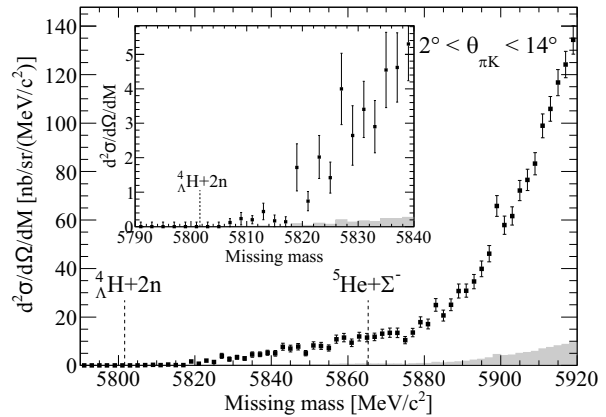}
  \caption{Missing-mass spectrum of the $^{6}$Li($\pi^{-}, K^{+}$)X reaction in $2^{\circ} < \theta_{\pi K} < 14^{\circ}$ angular range at 1.2-GeV/$c$ beam momentum. The vertical axis is the double-differential cross section averaged over the $\theta_{\pi K}$ region. A magnified view in the vicinity of the $^{6}_{\Lambda}$H bound state region is given in the inset plot. The error bars show the statistical errors. The systematic errors in each bin are represented by the bar graph at the bottom. The two vertical dashed lines represent the mass thresholds, $^{4}_{\Lambda}{\mathrm{H}}+2n$ and $^{5}{\mathrm{He}} + \Sigma^{-}$.}
  \label{fig:h6l-2-14}
\end{figure}

Figure~\ref{fig:h6l-2-14} shows the angular averaged spectrum in the angular range of $2^{\circ}<\theta_{\pi K}<14^{\circ}$.
In the figure the error bars show the statistical errors and the gray bar graph represents the systematic errors.
The horizontal and vertical axes correspond to the missing mass and to the double-differential cross section in units of nb/sr/(MeV/$c^{2}$), respectively.
The numerical values of the cross sections are listed in Appendix~\ref{AppB}, along with the statistical and systematic errors.
The two vertical dashed lines in the figure correspond to the mass thresholds, $^{4}_{\Lambda}{\mathrm{H}} +2n$ and $^{5}{\mathrm{He}} + \Sigma^{-}$.
It is apparent that the missing-mass count is zero below the $^{4}_{\Lambda}{\mathrm{H}} +2n$ threshold of 5801.6 MeV/$c^{2}$.
The mass of $^{4}_{\Lambda}{\mathrm{H}}$ from Ref.~\cite{H4L} was used in the threshold calculation.

\begin{figure}[htb]
  \includegraphics[width=8.5cm,clip]{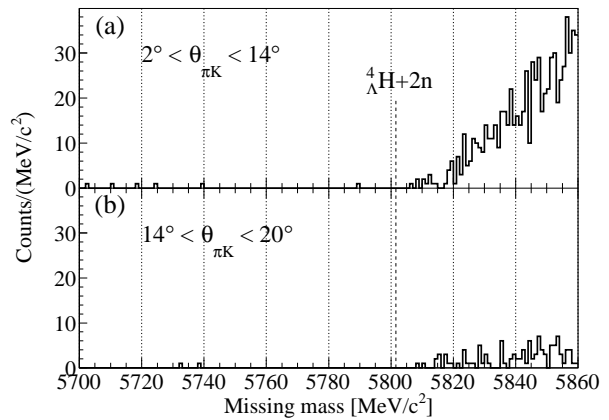}
  \caption{Raw missing-mass spectrum in angular range of (a) $2^{\circ}<\theta_{\pi K}<14^{\circ}$ and (b) $14^{\circ}<\theta_{\pi K}<20^{\circ}$.}
  \label{fig:h6l_count}
\end{figure}

Here, we estimated the number of background events expected in the vicinity of the bound region of $^{6}_{\Lambda}$H from the events lying between 5700--5790 MeV/$c^{2}$.
Figure~\ref{fig:h6l_count} (a) shows the raw missing-mass spectrum in the angular range of $2^{\circ} < \theta_{\pi K} < 14^{\circ}$. 
Six events can be found in the unphysical missing-mass region.
These events are interpreted as corresponding to $\pi^{+}$ or proton contamination, because there is no possible $K^{+}$ production process. 
By assuming that the degree of contamination does not strongly depend on the missing mass, we obtained an averaged background level of $0.060\pm 0.024$ event/(MeV/$c^{2}$).
Thus, 0.30 background counts were expected in the 2$\sigma$ region of the missing-mass resolution in the vicinity of the bound region, which is consistent with the null result.
On the other hand, three events were found in the bound region in the previous analysis \cite{Sugimura}.
By taking into account the background level of $0.39\pm0.05$ event/(MeV/$c^{2}$) in the previous analysis, it becomes apparent that these three events are also consistent with the expected number of background counts in the 2$\sigma$ region of the missing-mass resolution.
Thus, we classified these three events as background in the present analysis, owing to the improvement in the $K^{+}$ identification methods.

From the spectrum of Fig.~\ref{fig:h6l-2-14}, we estimated the upper limit for the $^{6}_{\Lambda}$H production cross section.
We assumed that the background-free hypothesis was valid in order to estimate the upper limit conservatively. 
Then, the count-base upper limit (90\% C.L.) for the null event was 2.3 events.
As the angular distribution of the production cross section was unknown, the correction factors used to obtain the upper limit in the cross section unit were evaluated as follows.
In order to estimate the expected analysis efficiency in the vicinity of 5800 MeV/$c^{2}$, we first calculated the correction factors in two-degree intervals in the $\Lambda$ continuum region using actual events, because several analysis efficiencies have an angular dependence.
Using the correction factors obtained in two-degree intervals, the average angular correction factor from 2 to 14 degrees was obtained and employed, assuming a flat production cross section distribution in the lab system.
Thus, the upper limit for the $^{6}_{\Lambda}$H production cross section was turned out to be 0.56 nb/sr (90\% C.L.).
The present upper limit is roughly 20 times smaller than the integrated cross section of 11.3$\:\pm\:$1.9 nb/sr obtained in the bound region of $^{10}_{\Lambda}$Li in the KEK-PS E521 experiment \cite{E521}.

The lowest states of $^{6}_{\Lambda}$H are assumed to be a spin doublet of $0^{+}$ and $1^{+}$ states, with the $0^{+}$ state being expected to be more deeply bound.
On the other hand, the ground state of $^{6}$Li has an almost pure $L=0$ and $S=1$ configuration \cite{Li6GS}.
Thus, spin-flip is necessary to populate the ground $0^{+}$ state directly via the ($\pi^{-}, K^{+}$) reaction.
However, the spin-flip amplitude is in general small in the ($\pi, K$) reaction.
Further, the spin-flip cross section is small for small forward reaction angles and the cross section increases with an increase in the reaction angle, as discussed in Ref.~\cite{FINUDA2}.
Thus, the $1^{+}$ state is dominantly populated at the forward reaction angle via the $^{6}$Li($\pi^{-}, K^{+}$)X reaction; however, the $0^{+}$ state is populated in the case of a large reaction angle only.
Thus, we investigated events with a reaction angle larger than 14 degrees. 
Figure~\ref{fig:h6l_count} (b) shows the raw missing-mass spectrum in the angular range of $14^{\circ} < \theta_{\pi K} < 20^{\circ}$.
However, there is no event in the vicinity of the $^{4}_{\Lambda}{\mathrm{H}} + 2n$ threshold in Fig.~\ref{fig:h6l_count} (b).

Although we discussed events below the $^{4}_{\Lambda}{\mathrm{H}}+ 2n$ threshold, no event was observed even above the threshold.
Then, we estimated the null event region in the present spectrum by considering the missing-mass resolution of 2.9 MeV (FWHM).
In the present discussion, the missing-mass scale error of 350 keV/$c^{2}$ was ignored, because this value is negligibly small compared with the missing-mass resolution.
The smallest missing-mass value of the $K^{+}$ event in Fig.~\ref{fig:h6l-2-14} was 5806.4 MeV/$c^{2}$, according to the present analysis.
We set a 2 MeV/$c^{2}$ margin from the last event, which fell in the 1.65$\sigma$ (90\% C.L.) range of the missing-mass resolution, and we excluded this from the null event region.
Therefore, there was no $K^{+}$ event up to 2.8 MeV/$c^{2}$, a value larger than the mass of $^{4}_{\Lambda}{\mathrm{H}}+2n$, even if we considered the missing-mass resolution.
In other words, the null event region in the present result extended up to 5804.4 MeV/$c^{2}$.

We compare our result with that reported in Ref.~\cite{FINUDA2}.
The FINUDA Collaboration reported evidence of the bound state of $^{6}_{\Lambda}$H.
In the FINUDA scenario, the $1^{+}$ state was populated by the $^{6}$Li($K^{-}_{\mathrm{stopped}}, \pi^{+}$) reaction and then it decayed to the $0^{+}$ state via $\gamma$-ray emission.
The $^{6}_{\Lambda}$H averaged mass from the production process reported by the FINUDA Collaboration was 5801.9 MeV \cite{FINUDA2}; however, in our result, no event was observed in the vicinity of this mass region.
In the present experiment, the $1^{+}$ state was not observed.
As mentioned above, the $1^{+}$ state is dominantly populated via the ($\pi^{-}, K^{+}$) reaction as well as the ($K^{-}_{\mathrm{stopped}}, \pi^{+}$) reaction.
Although the production reactions in the two experiments differed, it is not easy to explain the different results, because the momentum transfers of the ($\pi^{-}, K^{+}$) and the ($K^{-}_{\mathrm{stopped}}, \pi^{+}$) reactions are similar.

Finally, we discuss the relation between our result and recent theoretical predictions.
The existence of a bound state of $^{6}_{\Lambda}$H has been predicted by Akaishi and Yamazaki \cite{Akaishi_H6L} and by Gal and Millener \cite{Gal_H6L}.
Akaishi and Yamazaki predicted that the $^{6}_{\Lambda}$H hypernucleus was deeply bound, because of the strong $\Lambda$-$\Sigma$ mixing, while Gal and Millener estimated smaller $\Lambda$-$\Sigma$ mixing.
If the bound state is populated via the one-step reaction, the production cross section becomes larger as a result of the increase in the strength of the $\Lambda$-$\Sigma$ mixing.
Then, the theoretical expectation by Gal and Millener is preferred according to the interplay between the production cross section and the strength of the $\Lambda$-$\Sigma$ mixing due to the small production cross section upper limit obtained in the present experiment.

As Hiyama \textit{et~al.} \cite{Hiyama_H6L} suggested, even the $0^{+}$ state of $^{6}_{\Lambda}$H is unbound.
If the wave function of the core nucleus $^{5}$H is spatially broad, the system cannot gain sufficient attraction, because of the small overlap between the $\Lambda$ particle and the core nucleus.
The theoretical expectation indicates that the overlap between the $\Lambda$ and the core nucleus is sensitive to the mass and width of the $^{5}$H resonance state.
In fact, we can experimentally observe such a unbound ground state as the resonance peak in the vicinity of the bound threshold.
However, the strength of the $\Lambda$-$\Sigma$ mixing is related to the overlap between the $\Lambda$ and nucleons.
Thus, the spatial distribution of the core-nucleus wave function may be critical, because the small $\Lambda$-$\Sigma$ mixing causes the small production cross section in the case of the one-step DCX reaction.
The present result is in favor of such a scenario.

\subsection{Production cross section of $\Lambda$ continuum and $\Sigma$ quasi-free regions}
The increase in the cross section from the $^{5}{\mathrm{He}} +\Sigma^{-}$ threshold on ward in Fig.~\ref{fig:h6l-2-14} is due to the contribution of the $\Sigma^{-}$ quasi-free production process. 
On the other hand, events below the $^{5}{\mathrm{He}} + \Sigma^{-}$ threshold are recognized as the $\Lambda$ continuum region. 
In this experiment, the $\Lambda$ continuum and part of the $\Sigma^{-}$ quasi-free production region via the $^{6}$Li($\pi^{-}, K^{+}$)X reaction were simultaneously measured with high statistics.

In the KEK-PS E438 experiment \cite{E438}, spectrum fitting based on the Green's function method was performed to estimate the $\Sigma$-nucleus optical potential for the $\Sigma^{-}$-$^{27}$Al system \cite{Harada_E438}.
The present data allow us to investigate the $\Sigma$-nucleus optical potential for the $\Sigma^{-}$-$^{5}$He system more precisely, owing to the higher statistics compared to the KEK-PS E438 one.
The $\Lambda$ continuum is produced via the conversion of $\Sigma$ particles in a nucleus, that is its yield relates to the strength of the imaginary part of the $\Sigma$-nucleus optical potential.
On the other hand, the shape of the $\Sigma$ quasi-free region is affected by the real part of the potential.
If the $\Sigma$-nucleus optical potential is repulsive, the spectrum is enhanced in the higher-mass region and its increase at the threshold becomes slow.
In the KEK-PS E438 experiment, target nuclei with mass numbers greater than 12 were used.
As the isospin-dependent part of the $\Sigma$-nucleus optical potential is minor in heavy nuclei, only the spin-isospin averaged $\Sigma^{-}$-nucleus optical potential was investigated.
On the other hand, the $\Sigma$-nucleus optical potential depends on the isospin of the core nucleus $\Sigma^{-}$ in the light nucleus.
In particular, the $\Sigma^{-}$-$n$ interaction contribution was examined because the final state obtained via the DCX reaction was a neutron-rich environment.

\begin{turnpage}
\begin{figure*}[htb]
  \includegraphics[width=23.0cm,clip]{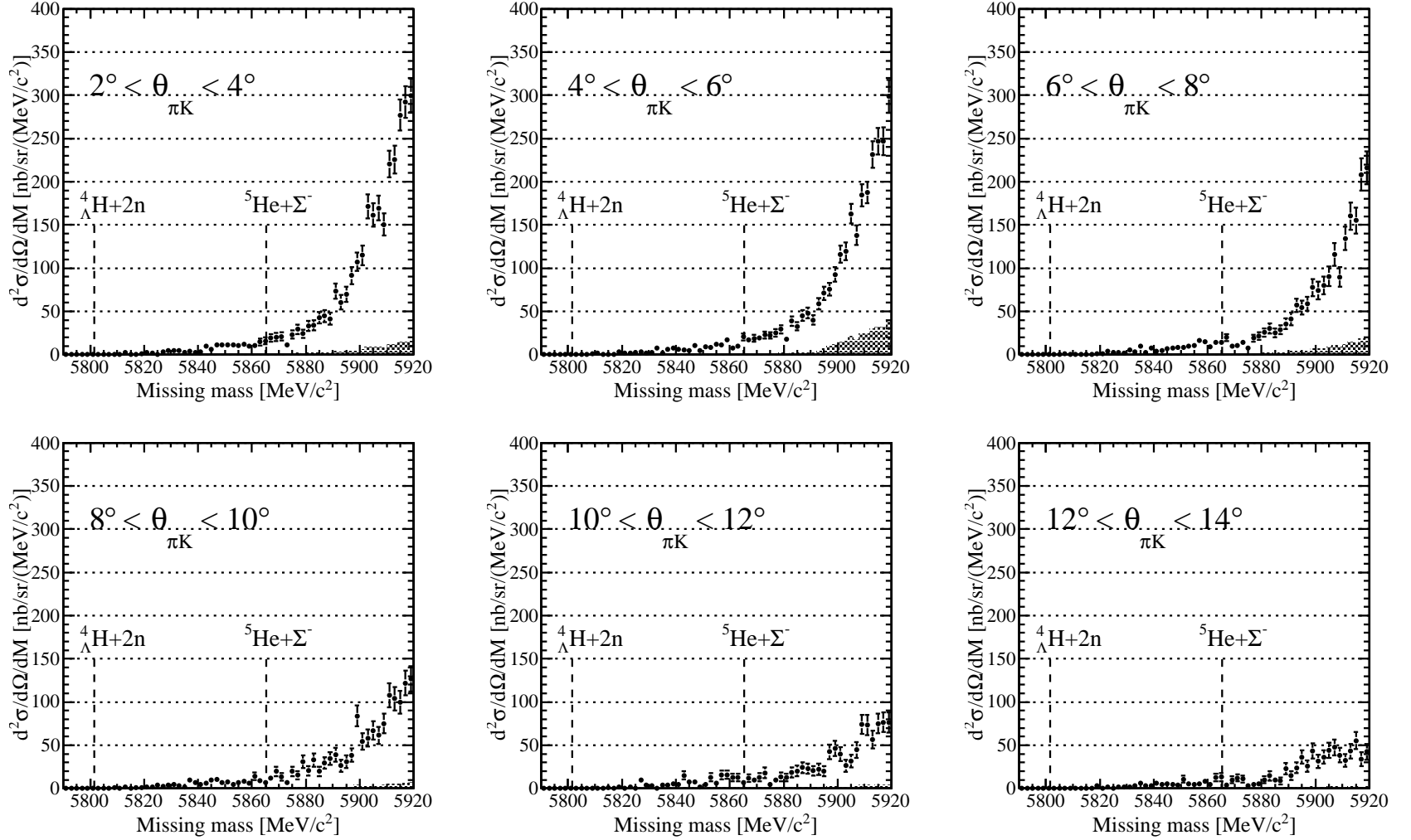}
  \caption{Missing-mass spectra of the $^{6}Li$($\pi^{-}, K^{+}$)X reaction in every two-degree bin width. The missing-mass spectra are shown in the same manner as in Fig.~\ref{fig:h6l-2-14}.}
  \label{fig:h6l-div}
\end{figure*}
\end{turnpage}

\begin{figure}[htb]
  \includegraphics[width=8.5cm,clip]{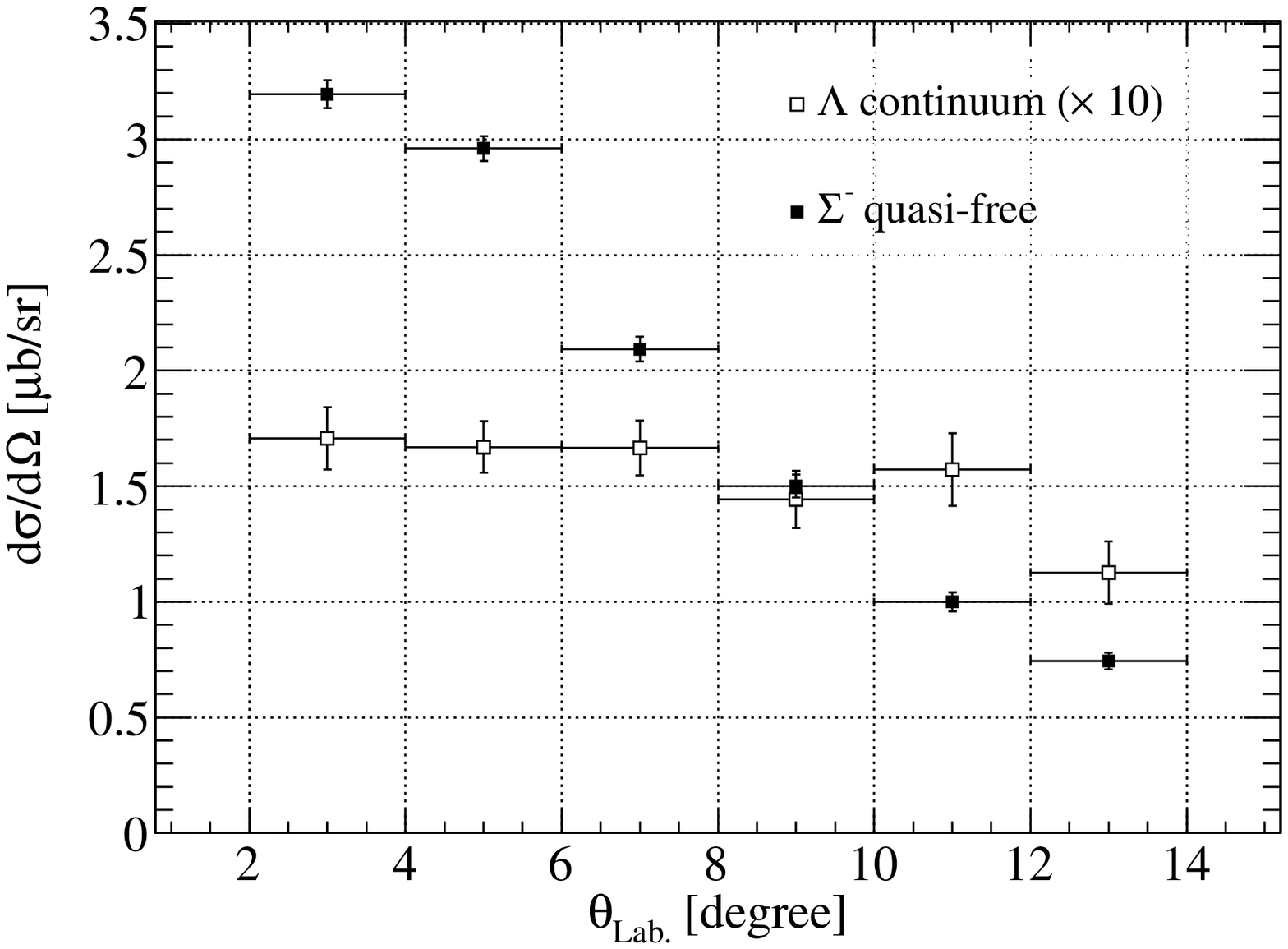}
  \caption{Angular distribution of the integrated cross sections of the $^{6}$Li($\pi^{-}, K^{+}$)X reaction. The closed squares represent the $\Sigma^{-}$ quasi-free production cross section integrated between the missing-mass range of 5865 to 5920 MeV/$c^{2}$. The open squares indicate the cross section of the $\Lambda$ continuum integrated between the missing-mass range of 5800 to 5865 MeV/$c^{2}$, multiplied by a factor of 10. The error bars are statistical only.}
  \label{fig:integ_h6l}
\end{figure}

As a result of the high statistics, we were able to obtain several spectra as a function of different reaction angles sub-ranges.
The $^{6}$Li($\pi^{-}, K^{+}$)X spectra obtained in each two-degree interval are shown in Fig.~\ref{fig:h6l-div}, in the same manner as Fig.~\ref{fig:h6l-2-14}.
The numerical values of the double-differential cross sections are summarized in Appendix~\ref{AppC}.
A clear difference in angular dependence is apparent between the $\Lambda$ continuum and the $\Sigma^{-}$ quasi-free production region.
The cross section at approximately 5920 MeV/$c^{2}$ in the angular region of $12^{\circ} < \theta_{\pi K} < 14^{\circ}$, is roughly a sixth of obtained for $2^{\circ} < \theta_{\pi K} < 4^{\circ}$.
On the other hand, the cross section below the $\mathrm{{}^{5}He} + \Sigma^{-}$ threshold is almost independent of the reaction angle.
The difference between the angular distributions of the $\Lambda$ continuum and of the $\Sigma^{-}$ quasi-free region becomes clear upon integration of the cross sections in both missing-mass regions, as shown in Fig.~\ref{fig:integ_h6l}.
The error bars represent statistical errors only.
The integration range of the missing mass is from 5800 to 5865 MeV/$c^{2}$ for the $\Lambda$ continuum and from 5865 to 5920 MeV/$c^{2}$ for the $\Sigma^{-}$ production reaction.
The results of the $\Lambda$ continuum are multiplied by a factor of 10 in Fig.~\ref{fig:integ_h6l}.

Several possible reasons for the different angular dependence can be considered.
As a $\Sigma^{-}$ particle was produced by the quasi-free reaction, it is not surprising that the angular distribution is similar to that of the $\pi^{-} p \rightarrow K^{+} \Sigma^{-}$ reaction obtained in Sec.~\ref{sec:elem}.
On the other hand, both one-step and two-step reactions can contribute to the $\Lambda$ continuum.
Thus, the flat angular distribution may be due the two-step reaction, which is independent of the $\pi^{-} p \rightarrow K^{+} \Sigma^{-}$ reaction.
In addition, the particle-hole state could affect the spectrum shape.
In the $\Sigma^{-}$ quasi-free region, a $\Sigma^{-}$ particle tends to be produced from a proton on the $^{6}$Li surface.
However, in the case of the $\Lambda$ continuum region, the probability of producing the hole state in the $s$-orbit may be higher, because two protons are involved in the reaction.
As the energy transfer differs for these two cases, the energy transfer dependence of the $\pi^{-} p \rightarrow K^{+} \Sigma^{-}$ reaction could appear in the angular distribution.
Finally, as the angular distribution of the ($\pi^{-}, K^{+}$) reaction was observed for the first time, further theoretical analysis is required in the near future.

\section{Summary}
We conducted the J-PARC E10 experiment, devoted to the search for the $^{6}_{\Lambda}$H neutron-rich hypernucleus.
This experiment was performed at the K1.8 beam line in the J-PARC Hadron Experimental Facility using the SKS complex as a kaon spectrometer.
A total of 1.4$\:\times\: 10^{12}$ $\pi^{-}$ beams at 1.2 GeV/$c$ were driven onto the $^{6}$Li target, which was 3.5-$\mathrm{g/cm^{2}}$ thickness and 95.54\% enriched.
In addition, ($\pi^{\pm}, K^{+}$) and ($\pi^{+}, K^{+}$) data were obtained using the polyethylene and graphite targets, respectively.
We employed an improved $K^{+}$ identification method in the present analysis.
As the result, a background level of $0.060 \pm 0.024$ events/(MeV/$c^{2}$) was achieved, significantly lower than that of the previous analysis.
Furthermore, the number of studied $K^{+}$ events was increased by 20\% compared to the previous analysis, by improving the analysis efficiencies.

Finally, we searched for the bound state of $^{6}_{\Lambda}$H in the missing-mass spectrum of the $^{6}$Li($\pi^{-}, K^{+}$)X reaction in the angular range of $2^{\circ}<\theta_{\pi K}<14^{\circ}$ with a missing-mass resolution of 2.9 MeV/$c^{2}$; however, no $K^{+}$ event was observed up to 2.8 MeV/$c^{2}$ above the mass threshold of $^{4}_{\Lambda}{\mathrm{H}}+2n$.
This means that neither the $0^{+}$ nor the $1^{+}$ state reported by the FINUDA experiment was observed.

The $\Lambda$ continuum region and part of the $\Sigma^{-}$ quasi-free production region via the $^{6}$Li($\pi^{-}, K^{+}$)X reaction were simultaneously measured with high statistics for the first time.
The present result may provide an opportunity to investigate the $\Sigma^{-}$-nucleus optical potential for the $\Sigma^{-}$-$^{5}$He system precisely via spectrum fitting.
In addition, the $^{6}$Li($\pi^{-}, K^{+}$)X spectra were obtained for two-degree intervals.
It was found that the angular distribution of the $\Sigma^{-}$ quasi-free region was steep, while that of the $\Lambda$ continuum region was almost independent of the reaction angle.

\acknowledgments
We would like to acknowledge the staff of the J-PARC accelerator and the Hadron Experimental Facility.
The present experiment would not be possible without their efforts.
This work was supported by various Grants-in-Aid for Scientiﬁc Research (KAKENHI): Scientiﬁc Research on Innovative Areas (No.~JP24105003), Young Scientist (A) (No.~JP23684011), and Scientific Research (C) (No.~JP24540305).
This work was also supported by a Basic Research (Young Researcher) (No.~2010-0004752) award from the National Research Foundation of Korea. 
We would like to acknowledge support from WCU program, National Research Foundation, Center for Korean J-PARC Users, and the Ministry of Education, Science, and Technology (Korea).

\appendix
\section{}
\label{AppA}
The $\Sigma^{\pm}$ production cross sections obtained via the $\pi^{\pm} p \rightarrow K^{+} \Sigma^{\pm}$ reactions at a beam momentum of 1.39 GeV/$c$ in the lab and C.M. systems are listed in Tables~\ref{tb:sigma_lab} and \ref{tb:sigma_cm}, respectively.

\begin{table}[htb]
  \caption{Summary of the $\Sigma^{\pm}$ production cross sections via the $\pi^{\pm} p \rightarrow K^{+} + \Sigma^{\pm}$ reactions at 1.39 GeV/$c$ in lab system.}
  \begin{tabular*}{8.5cm}{@{\extracolsep{\fill}}cccc} \hline \hline
    $\theta_{\mathrm{lab.}}$  & $d\sigma/d\Omega$  & \multicolumn{2}{c}{Errors} \\
                           &                    & stat.             & syst.     \\
    (deg)                  & ($\mu$b/sr)        & ($\mu$b/sr)       & ($\mu$b/sr) \\ \hline
    \multicolumn{4}{c}{$\Sigma^{-}$} \\ \hline
    $3^{\circ}$            & 79.66              & 2.63              & 9.40 \\ 
    $5^{\circ}$            & 80.85              & 2.25              & 9.22 \\ 
    $7^{\circ}$            & 71.91              & 2.25              & 8.34 \\ 
    $9^{\circ}$            & 56.66              & 2.03              & 6.46 \\ 
    $11^{\circ}$           & 48.34              & 2.10              & 5.51 \\ 
    $13^{\circ}$           & 38.83              & 2.31              & 4.54 \\ 
    $15^{\circ}$           & 26.95              & 2.02              & 3.13 \\ \hline
    \multicolumn{4}{c}{$\Sigma^{+}$} \\ \hline
    $3^{\circ}$            & 566.2              & 28.5              & 28.9 \\ 
    $5^{\circ}$            & 496.1              & 23.3              & 21.8 \\ 
    $7^{\circ}$            & 404.1              & 22.0              & 19.4 \\ 
    $9^{\circ}$            & 346.1              & 21.1              & 15.6 \\ 
    $11^{\circ}$           & 280.3              & 20.6              & 20.2 \\ 
    $13^{\circ}$           & 228.6              & 23.9              & 18.5 \\ 
    $15^{\circ}$           & 159.9              & 21.6              & 18.6 \\ \hline \hline
  \end{tabular*}
  \label{tb:sigma_lab}
\end{table}

\begin{table}[htb]
  \caption{Summary of the $\Sigma^{\pm}$ production cross sections via the $\pi^{\pm} p \rightarrow K^{+} + \Sigma^{\pm}$ reactions at 1.39 GeV/$c$ in C.M. system.}
  \begin{tabular*}{8.5cm}{@{\extracolsep{\fill}}cccc} \hline \hline
    cos$\theta_{\mathrm{C.M.}}$& $d\sigma/d\Omega$ & \multicolumn{2}{c}{Errors} \\
                           &                    & stat.             & syst.     \\
    (deg)                  & ($\mu$b/sr)        & ($\mu$b/sr)       & ($\mu$b/sr) \\ \hline
    \multicolumn{4}{c}{$\Sigma^{-}$} \\ \hline
    0.975                  & 12.91              & 0.25              & 1.72 \\
    0.925                  & 9.96               & 0.28              & 1.40 \\
    0.875                  & 7.47               & 0.35              & 1.20 \\
    0.825                  & 5.94               & 0.38              & 1.06 \\ \hline
    \multicolumn{4}{c}{$\Sigma^{+}$} \\ \hline
    0.975                  & 80.69              & 2.55              & 3.55 \\
    0.925                  & 53.10              & 2.81              & 2.39 \\
    0.875                  & 46.23              & 3.59              & 3.51 \\
    0.825                  & 36.06              & 4.01              & 3.00 \\ \hline \hline
  \end{tabular*}
  \label{tb:sigma_cm}
\end{table}

\section{}
\label{AppB}
We summarize the double-differential cross secitons of the $^{6}_{\Lambda}$Li($\pi^{-}, K^{+}$)X reaction in the angular range of $2^{\circ}<\theta_{\pi K}<14^{\circ}$ in Table \ref{tb:h6l-all}.
\LTcapwidth=\columnwidth
\begin{longtable}[htb]{@{\extracolsep{\fill}}cccc}
  \caption{Angular averaged cross sections of the $^{6}_{\Lambda}$Li($\pi^{-}, K^{+}$)X reaction between 2 and 14 degrees.}\\\hline\hline
  \label{tb:h6l-all}
  \endfirsthead
  \caption*{TABLE \ref{tb:h6l-all}: (Continued.)}\\\hline\hline
  \endhead
  \endfoot
  \hline\hline
  \endlastfoot
    Mass                   & $d^{2}\sigma/d\Omega/dM$  & \multicolumn{2}{c}{Errors} \\
    (MeV/$c^{2}$)          & (nb/sr/(MeV/$c^{2}$))     & stat.            & syst.     \\ \hline
    5791  &        0.00  &        0.13  &        0.00 \\
    5793  &        0.00  &        0.13  &        0.00 \\ 
    5795  &        0.00  &        0.13  &        0.00 \\ 
    5797  &        0.00  &        0.13  &        0.00 \\ 
    5799  &        0.00  &        0.13  &        0.00 \\ 
    5801  &        0.00  &        0.13  &        0.00 \\ 
    5803  &        0.00  &        0.13  &        0.00 \\ 
    5805  &        0.00  &        0.13  &        0.00 \\ 
    5807  &        0.12  &        0.12  &        0.00 \\ 
    5809  &        0.24  &        0.18  &        0.00 \\ 
    5811  &        0.21  &        0.12  &        0.00 \\ 
    5813  &        0.44  &        0.24  &        0.02 \\ 
    5815  &        0.17  &        0.17  &        0.00 \\ 
    5817  &        0.58  &        0.46  &        0.00 \\ 
    5819  &        1.79  &        0.69  &        0.09 \\ 
    5821  &        0.92  &        0.33  &        0.05 \\ 
    5823  &        1.75  &        0.58  &        0.09 \\ 
    5825  &        2.08  &        0.67  &        0.00 \\ 
    5827  &        4.02  &        1.04  &        0.21 \\ 
    5829  &        2.51  &        0.82  &        0.13 \\ 
    5831  &        3.82  &        0.91  &        0.20 \\ 
    5833  &        2.72  &        0.68  &        0.14 \\ 
    5835  &        5.21  &        1.17  &        0.27 \\ 
    5837  &        4.60  &        1.01  &        0.24 \\ 
    5839  &        5.07  &        1.05  &        0.26 \\ 
    5841  &        4.75  &        1.10  &        0.24 \\ 
    5843  &        7.21  &        1.38  &        0.37 \\ 
    5845  &        8.65  &        1.62  &        0.44 \\ 
    5847  &        8.52  &        1.43  &        0.44 \\ 
    5849  &        5.96  &        1.22  &        0.31 \\ 
    5851  &        8.26  &        1.41  &        0.42 \\ 
    5853  &        7.77  &        1.45  &        0.40 \\ 
    5855  &        8.17  &        1.40  &        0.42 \\ 
    5857  &       11.06  &        1.66  &        0.57 \\ 
    5859  &       12.13  &        1.81  &        0.62 \\ 
    5861  &        9.74  &        1.61  &        0.50 \\ 
    5863  &       12.77  &        1.91  &        0.66 \\ 
    5865  &       11.82  &        1.77  &        0.61 \\ 
    5867  &       14.20  &        1.89  &        0.73 \\ 
    5869  &       13.94  &        1.88  &        0.72 \\ 
    5871  &       14.27  &        1.98  &        0.73 \\ 
    5873  &       14.38  &        2.09  &        0.74 \\
    5875  &       10.38  &        1.48  &        0.53 \\ 
    5877  &       15.97  &        1.94  &        0.82 \\ 
    5879  &       19.03  &        2.28  &        0.98 \\ 
    5881  &       17.77  &        2.08  &        0.91 \\ 
    5883  &       25.81  &        2.62  &        1.32 \\ 
    5885  &       23.15  &        2.53  &        1.19 \\ 
    5887  &       26.33  &        2.58  &        1.35 \\ 
    5889  &       32.86  &        3.11  &        1.69 \\ 
    5891  &       30.46  &        2.71  &        1.56 \\ 
    5893  &       35.89  &        3.07  &        1.84 \\ 
    5895  &       42.83  &        3.44  &        2.20 \\ 
    5897  &       47.88  &        3.49  &        2.46 \\ 
    5899  &       65.33  &        4.17  &        3.35 \\ 
    5901  &       63.24  &        4.08  &        3.24 \\ 
    5903  &       65.17  &        4.01  &        3.34 \\ 
    5905  &       76.34  &        4.31  &        3.92 \\ 
    5907  &       77.55  &        4.47  &        3.98 \\ 
    5909  &       88.81  &        4.82  &        4.56 \\ 
    5911  &      102.17  &        5.07  &        5.24 \\ 
    5913  &      107.63  &        5.16  &        5.52 \\ 
    5915  &      121.01  &        5.64  &        6.21 \\ 
    5917  &      123.43  &        5.47  &        6.33 \\ 
    5919  &      143.63  &        6.10  &        7.37 \\ 
    5921  &      155.42  &        6.52  &        7.97 \\
\end{longtable}

\section{}
\label{AppC}
We summarize the double-differential cross secitons of the $^{6}_{\Lambda}$Li($\pi^{-}, K^{+}$)X reaction in two-degree intervals in Tables \ref{tb:h6l-div1}, \ref{tb:h6l-div2}, and \ref{tb:h6l-div3}.

\LTcapwidth=2\columnwidth

\begin{longtable*}[H]{@{\extracolsep{\fill}}ccccccc}
  \caption{Cross sections of the $^{6}_{\Lambda}$Li($\pi^{-}, K^{+}$)X reaction between 2 and 4 degrees and between 4 and 6 degrees.}\\\hline\hline
  \label{tb:h6l-div1}
  \endfirsthead
  \caption*{TABLE \ref{tb:h6l-div1}: (Continued.)}\\\hline\hline
  \endhead
  \endfoot
  \hline\hline
  \endlastfoot
    Mass            & $d^{2}\sigma/d\Omega/dM$  &\multicolumn{2}{c}{Errors} & $d^{2}\sigma/d\Omega/dM$  & \multicolumn{2}{c}{Errors} \\
    (MeV/$c^{2}$)   & (nb/sr/(MeV/$c^{2}$))     & stat.        & syst.      & (nb/sr/(MeV/$c^{2}$))     & stat.        & syst. \\
                    & 2--4 deg                  &              &            & 4--6 deg                   &              &      \\ \hline
    5791  &        0.00  &        1.17  &        0.00   &        0.00  &        0.81  &        0.00   \\  
    5793  &        0.00  &        1.17  &        0.00   &        0.00  &        0.81  &        0.00   \\  
    5795  &        0.00  &        1.17  &        0.00   &        0.00  &        0.81  &        0.00   \\  
    5797  &        0.00  &        1.17  &        0.00   &        0.00  &        0.81  &        0.00   \\  
    5799  &        0.00  &        1.17  &        0.00   &        0.00  &        0.81  &        0.00   \\  
    5801  &        0.00  &        1.17  &        0.00   &        0.00  &        0.81  &        0.00   \\  
    5803  &        0.00  &        1.17  &        0.00   &        0.00  &        0.81  &        0.00   \\  
    5805  &        0.00  &        1.17  &        0.00   &        0.00  &        0.81  &        0.00   \\  
    5807  &        0.00  &        1.17  &        0.00   &        0.00  &        0.81  &        0.00   \\  
    5809  &        0.00  &        1.17  &        0.00   &        0.64  &        0.64  &        0.03   \\  
    5811  &        0.00  &        1.17  &        0.00   &        1.99  &        1.15  &        0.10   \\  
    5813  &        2.14  &        1.51  &        0.11   &        0.00  &        0.81  &        0.00   \\  
    5815  &        0.00  &        1.17  &        0.00   &        0.00  &        0.81  &        0.00   \\  
    5817  &        0.00  &        1.17  &        0.00   &        0.00  &        0.81  &        0.00   \\  
    5819  &        1.00  &        1.00  &        0.05   &        3.69  &        1.66  &        0.19   \\  
    5821  &        2.58  &        1.89  &        0.13   &        1.86  &        1.08  &        0.09   \\  
    5823  &        1.77  &        1.25  &        0.09   &        1.22  &        0.86  &        0.06   \\  
    5825  &        0.00  &        1.17  &        0.00   &        2.45  &        1.51  &        0.12   \\  
    5827  &        3.22  &        1.86  &        0.17   &        3.07  &        1.54  &        0.16   \\  
    5829  &        4.84  &        2.17  &        0.25   &        1.52  &        1.08  &        0.08   \\  
    5831  &        5.96  &        2.44  &        0.31   &        2.13  &        1.25  &        0.11   \\  
    5833  &        4.92  &        2.20  &        0.25   &        8.74  &        2.56  &        0.45   \\  
    5835  &        3.74  &        1.87  &        0.19   &        2.07  &        1.22  &        0.11   \\  
    5837  &        3.89  &        1.95  &        0.20   &        5.12  &        1.98  &        0.26   \\  
    5839  &        1.22  &        1.22  &        0.06   &        8.04  &        2.35  &        0.41   \\  
    5841  &        3.06  &        1.77  &        0.16   &        5.90  &        1.98  &        0.30   \\  
    5843  &        9.61  &        3.28  &        0.49   &        5.70  &        2.08  &        0.29   \\  
    5845  &        6.51  &        3.01  &        0.33   &        4.99  &        1.78  &        0.25   \\  
    5847  &       10.48  &        3.38  &        0.54   &       10.39  &        2.72  &        0.53   \\  
    5849  &       12.78  &        3.69  &        0.66   &        6.37  &        2.15  &        0.32   \\  
    5851  &       13.19  &        3.86  &        0.68   &       11.62  &        3.27  &        0.59   \\  
    5853  &        9.83  &        3.34  &        0.50   &        7.74  &        2.36  &        0.39   \\  
    5855  &       10.31  &        3.27  &        0.53   &       11.90  &        3.02  &        0.61   \\  
    5857  &       12.26  &        3.82  &        0.63   &       11.48  &        2.90  &        0.59   \\  
    5859  &        9.93  &        3.15  &        0.51   &       16.46  &        3.85  &        0.84   \\  
    5861  &       10.87  &        3.29  &        0.56   &        7.64  &        2.47  &        0.39   \\  
    5863  &       14.05  &        4.03  &        0.72   &        8.96  &        2.62  &        0.46   \\  
    5865  &       16.95  &        4.25  &        0.87   &       20.17  &        4.06  &        1.03   \\  
    5867  &       20.55  &        4.61  &        1.05   &       19.86  &        4.21  &        1.01   \\  
    5869  &       21.38  &        5.05  &        1.10   &       19.01  &        4.01  &        0.97   \\  
    5871  &       20.75  &        4.66  &        1.06   &       17.67  &        3.56  &        0.90   \\  
    5873  &       12.30  &        3.57  &        0.63   &       22.54  &        4.13  &        1.15   \\  
    5875  &       23.74  &        5.14  &        1.22   &       21.07  &        3.89  &        1.07   \\  
    5877  &       29.70  &        5.67  &        1.52   &       27.63  &        4.76  &        1.41   \\  
    5879  &       23.62  &        4.94  &        1.21   &       31.29  &        4.92  &        1.60   \\  
    5881  &       36.71  &        6.26  &        1.88   &       17.09  &        3.61  &        0.87   \\  
    5883  &       37.57  &        6.28  &        1.93   &       39.75  &        5.57  &        2.02   \\  
    5885  &       43.87  &        6.88  &        2.25   &       33.47  &        5.11  &        1.70   \\  
    5887  &       44.82  &        6.86  &        2.30   &       43.20  &        5.89  &        2.20   \\  
    5889  &       46.25  &        7.30  &        2.37   &       50.30  &        6.30  &        2.56   \\  
    5891  &       72.22  &        8.83  &        3.71   &       41.94  &        5.75  &        2.14   \\  
    5893  &       59.42  &        8.09  &        3.05   &       60.64  &        6.87  &        5.70   \\  
    5895  &       71.31  &        8.90  &        3.66   &       69.43  &        7.45  &        6.52   \\  
    5897  &       95.14  &       10.27  &        4.88   &       77.38  &        7.81  &       10.02   \\  
    5899  &      113.26  &       11.14  &        5.81   &       93.80  &        8.47  &       12.14   \\  
    5901  &      115.98  &       11.47  &        5.95   &      120.04  &       10.02  &       15.54   \\  
    5903  &      176.00  &       14.25  &        9.03   &      121.49  &       10.19  &       15.73   \\  
    5905  &      166.73  &       13.85  &        8.55   &      166.00  &       11.62  &       21.49   \\  
    5907  &      177.01  &       14.51  &        9.08   &      142.74  &       11.47  &       18.48   \\  
    5909  &      151.21  &       12.97  &        7.76   &      182.13  &       12.66  &       23.57   \\  
    5911  &      230.47  &       16.07  &       11.82   &      199.68  &       13.28  &       25.85   \\  
    5913  &      229.65  &       16.28  &       11.78   &      238.31  &       15.79  &       30.84   \\  
    5915  &      286.52  &       18.66  &       14.70   &      265.41  &       18.07  &       34.35   \\  
    5917  &      298.76  &       18.85  &       15.33   &      253.42  &       16.58  &       32.80   \\  
    5919  &      306.95  &       19.58  &       15.75   &      310.62  &       19.48  &       40.21   \\  
    5921  &      351.72  &       21.61  &       18.04   &      329.20  &       21.16  &       42.62   \\
\end{longtable*}

\begin{longtable*}[H]{@{\extracolsep{\fill}}ccccccc}
  \caption{Cross sections of the $^{6}_{\Lambda}$Li($\pi^{-}, K^{+}$)X reaction between 6 and 8 degrees and between 8 and 10 degrees.}\\\hline\hline
  \label{tb:h6l-div2}
  \endfirsthead
  \caption*{TABLE \ref{tb:h6l-div2}: (Continued.)}\\\hline\hline
  \endhead
  \endfoot
  \hline\hline
  \endlastfoot
    Mass            & $d^{2}\sigma/d\Omega/dM$  &\multicolumn{2}{c}{Errors} & $d^{2}\sigma/d\Omega/dM$  & \multicolumn{2}{c}{Errors} \\
    (MeV/$c^{2}$)   & (nb/sr/(MeV/$c^{2}$))     & stat.        & syst.      & (nb/sr/(MeV/$c^{2}$))     & stat.        & syst. \\
                    & 6--8 deg                  &              &            & 8--10 deg                 &              &      \\ \hline
    5791  &        0.00  &        0.94  &        0.00   &        0.00  &        1.01  &        0.00   \\  
    5793  &        0.00  &        0.94  &        0.00   &        0.00  &        1.01  &        0.00   \\  
    5795  &        0.00  &        0.94  &        0.00   &        0.00  &        1.01  &        0.00   \\  
    5797  &        0.00  &        0.94  &        0.00   &        0.00  &        1.01  &        0.00   \\  
    5799  &        0.00  &        0.94  &        0.00   &        0.00  &        1.01  &        0.00   \\  
    5801  &        0.00  &        0.94  &        0.00   &        0.00  &        1.01  &        0.00   \\  
    5803  &        0.00  &        0.94  &        0.00   &        0.00  &        1.01  &        0.00   \\  
    5805  &        0.00  &        0.94  &        0.00   &        0.00  &        1.01  &        0.00   \\  
    5807  &        0.79  &        0.79  &        0.04   &        0.00  &        1.01  &        0.00   \\  
    5809  &        0.00  &        0.94  &        0.00   &        0.91  &        0.91  &        0.05   \\  
    5811  &        0.00  &        0.94  &        0.00   &        0.00  &        1.01  &        0.00   \\  
    5813  &        0.00  &        0.94  &        0.00   &        1.64  &        1.16  &        0.08   \\  
    5815  &        0.00  &        0.94  &        0.00   &        0.00  &        1.01  &        0.00   \\  
    5817  &        0.00  &        0.94  &        0.00   &        0.76  &        0.76  &        0.04   \\  
    5819  &        0.67  &        0.67  &        0.04   &        0.72  &        0.72  &        0.04   \\  
    5821  &        0.87  &        0.87  &        0.05   &        2.35  &        1.36  &        0.12   \\  
    5823  &        3.93  &        1.77  &        0.20   &        1.85  &        1.31  &        0.10   \\  
    5825  &        3.51  &        1.79  &        0.18   &        3.49  &        1.76  &        0.18   \\  
    5827  &        3.52  &        1.76  &        0.18   &        1.61  &        1.14  &        0.08   \\  
    5829  &        3.14  &        1.58  &        0.16   &        2.28  &        1.32  &        0.12   \\  
    5831  &        7.40  &        2.83  &        0.37   &        5.15  &        2.11  &        0.26   \\  
    5833  &        2.32  &        1.35  &        0.12   &        3.77  &        1.94  &        0.19   \\  
    5835  &        9.76  &        2.85  &        0.49   &        3.47  &        1.74  &        0.18   \\  
    5837  &        3.07  &        1.54  &        0.16   &        9.92  &        3.07  &        0.51   \\  
    5839  &        6.66  &        2.36  &        0.34   &        7.09  &        2.51  &        0.36   \\  
    5841  &        3.83  &        1.72  &        0.19   &        4.96  &        2.03  &        0.25   \\  
    5843  &        4.95  &        2.04  &        0.25   &        6.85  &        2.67  &        0.35   \\  
    5845  &        6.75  &        2.57  &        0.35   &        8.94  &        3.22  &        0.46   \\  
    5847  &        8.16  &        2.80  &        0.42   &       11.39  &        3.45  &        0.58   \\  
    5849  &        8.11  &        2.59  &        0.41   &        5.91  &        2.74  &        0.30   \\  
    5851  &        7.56  &        2.54  &        0.38   &        8.81  &        3.11  &        0.45   \\  
    5853  &        9.87  &        2.87  &        0.50   &        3.62  &        2.22  &        0.18   \\  
    5855  &        9.49  &        2.88  &        0.48   &        7.47  &        2.71  &        0.38   \\  
    5857  &       17.30  &        3.98  &        0.88   &        9.53  &        3.15  &        0.49   \\  
    5859  &       15.77  &        3.76  &        0.80   &        7.92  &        2.95  &        0.41   \\  
    5861  &        8.83  &        2.67  &        0.45   &       16.80  &        4.92  &        0.86   \\  
    5863  &       13.34  &        3.41  &        0.68   &       10.78  &        3.87  &        0.55   \\  
    5865  &       14.70  &        3.58  &        0.74   &        4.75  &        2.23  &        0.24   \\  
    5867  &       17.99  &        3.87  &        0.91   &       11.49  &        3.65  &        0.59   \\  
    5869  &       14.89  &        3.54  &        0.75   &       21.20  &        5.80  &        1.08   \\  
    5871  &       10.45  &        2.93  &        0.53   &       12.61  &        3.99  &        0.65   \\  
    5873  &       13.27  &        3.71  &        0.67   &        7.12  &        3.10  &        0.37   \\  
    5875  &        8.42  &        2.68  &        0.43   &       16.41  &        4.77  &        0.84   \\  
    5877  &       18.47  &        3.98  &        0.94   &       21.22  &        5.77  &        1.08   \\  
    5879  &       21.01  &        4.57  &        1.06   &       32.90  &        7.44  &        1.67   \\  
    5881  &       28.74  &        5.05  &        1.45   &       18.37  &        5.38  &        0.94   \\  
    5883  &       31.03  &        5.67  &        1.57   &       32.08  &        7.35  &        1.63   \\  
    5885  &       28.91  &        5.44  &        1.46   &       23.25  &        6.25  &        1.18   \\  
    5887  &       27.92  &        5.38  &        1.42   &       29.64  &        6.72  &        1.51   \\  
    5889  &       41.48  &        6.70  &        2.10   &       33.72  &        7.46  &        1.72   \\  
    5891  &       41.59  &        7.37  &        3.68   &       40.16  &        8.13  &        2.04   \\  
    5893  &       54.13  &        7.61  &        4.79   &       28.24  &        7.19  &        1.44   \\  
    5895  &       61.30  &        8.68  &        5.42   &       27.49  &        6.75  &        1.41   \\  
    5897  &       58.04  &        8.37  &        5.13   &       37.64  &        7.68  &        1.92   \\  
    5899  &       82.66  &       10.05  &        7.31   &       86.00  &       12.33  &        4.38   \\  
    5901  &       81.82  &       10.46  &        7.23   &       59.08  &       10.12  &        3.01   \\  
    5903  &       83.95  &       10.35  &        7.42   &       61.79  &       10.46  &        3.15   \\  
    5905  &       93.47  &       11.67  &        8.27   &       72.79  &       10.91  &        3.71   \\  
    5907  &      111.12  &       12.88  &        9.82   &       59.21  &        9.83  &        3.02   \\  
    5909  &      101.89  &       12.11  &        9.01   &       87.63  &       12.08  &        4.46   \\  
    5911  &      132.74  &       13.71  &       11.74   &      110.97  &       14.02  &        5.65   \\  
    5913  &      163.91  &       16.31  &       14.50   &      107.65  &       13.85  &        5.48   \\  
    5915  &      146.02  &       14.81  &       12.91   &      102.58  &       13.41  &        5.22   \\  
    5917  &      215.26  &       18.91  &       19.03   &      129.30  &       15.04  &        6.58   \\  
    5919  &      232.60  &       19.82  &       20.57   &      141.62  &       15.73  &        7.21   \\  
    5921  &      226.25  &       20.19  &       20.00   &      133.52  &       15.31  &        6.80   \\
\end{longtable*}

\begin{longtable*}[H]{@{\extracolsep{\fill}}ccccccc}
  \caption{Cross sections of the $^{6}_{\Lambda}$Li($\pi^{-}, K^{+}$)X reaction between 10 and 12 degrees and between 12 and 14 degrees.}\\\hline\hline
  \label{tb:h6l-div3}
  \endfirsthead
  \caption*{TABLE \ref{tb:h6l-div3}: (Continued.)}\\\hline\hline
  \endhead
  \endfoot
  \hline\hline
  \endlastfoot
    Mass            & $d^{2}\sigma/d\Omega/dM$  &\multicolumn{2}{c}{Errors} & $d^{2}\sigma/d\Omega/dM$  & \multicolumn{2}{c}{Errors} \\
    (MeV/$c^{2}$)   & (nb/sr/(MeV/$c^{2}$))     & stat.        & syst.      & (nb/sr/(MeV/$c^{2}$))     & stat.        & syst. \\
                    & 10--12 deg                &              &            & 12--14 deg                 &              &      \\ \hline
    5791  &        0.00  &        1.62  &        0.00   &        0.00  &        1.79  &        0.00   \\  
    5793  &        0.00  &        1.62  &        0.00   &        0.00  &        1.79  &        0.00   \\  
    5795  &        0.00  &        1.62  &        0.00   &        0.00  &        1.79  &        0.00   \\  
    5797  &        0.00  &        1.62  &        0.00   &        0.00  &        1.79  &        0.00   \\  
    5799  &        0.00  &        1.62  &        0.00   &        0.00  &        1.79  &        0.00   \\  
    5801  &        0.00  &        1.62  &        0.00   &        0.00  &        1.79  &        0.00   \\  
    5803  &        0.00  &        1.62  &        0.00   &        0.00  &        1.79  &        0.00   \\  
    5805  &        0.00  &        1.62  &        0.00   &        0.00  &        1.79  &        0.00   \\  
    5807  &        0.00  &        1.62  &        0.00   &        0.00  &        1.79  &        0.00   \\  
    5809  &        0.00  &        1.62  &        0.00   &        0.00  &        1.79  &        0.00   \\  
    5811  &        0.00  &        1.62  &        0.00   &        0.00  &        1.79  &        0.00   \\  
    5813  &        0.00  &        1.62  &        0.00   &        0.00  &        1.79  &        0.00   \\  
    5815  &        0.74  &        0.74  &        0.04   &        0.00  &        1.79  &        0.00   \\  
    5817  &        0.00  &        1.62  &        0.00   &        1.64  &        1.64  &        0.08   \\  
    5819  &        1.54  &        1.54  &        0.08   &        2.79  &        1.97  &        0.14   \\  
    5821  &        0.00  &        1.62  &        0.00   &        0.00  &        1.79  &        0.00   \\  
    5823  &        0.83  &        0.83  &        0.04   &        1.49  &        1.49  &        0.08   \\  
    5825  &        0.82  &        0.82  &        0.04   &        1.72  &        1.72  &        0.09   \\  
    5827  &       10.04  &        3.85  &        0.51   &        1.42  &        1.42  &        0.07   \\  
    5829  &        3.36  &        2.64  &        0.17   &        1.45  &        1.45  &        0.07   \\  
    5831  &        1.48  &        1.48  &        0.07   &        3.11  &        2.20  &        0.16   \\  
    5833  &        0.00  &        1.62  &        0.00   &        1.68  &        1.68  &        0.09   \\  
    5835  &        4.46  &        2.59  &        0.23   &        6.17  &        3.09  &        0.31   \\  
    5837  &        4.67  &        2.70  &        0.24   &        1.62  &        1.62  &        0.08   \\  
    5839  &        6.24  &        3.13  &        0.31   &        1.57  &        1.57  &        0.08   \\  
    5841  &        3.25  &        2.00  &        0.16   &        6.33  &        3.17  &        0.32   \\  
    5843  &       11.91  &        4.22  &        0.60   &        4.71  &        2.72  &        0.24   \\  
    5845  &       16.54  &        5.38  &        0.83   &        4.72  &        2.73  &        0.24   \\  
    5847  &        7.57  &        3.40  &        0.38   &        6.34  &        3.18  &        0.32   \\  
    5849  &        1.77  &        1.77  &        0.09   &        6.65  &        3.33  &        0.34   \\  
    5851  &        4.46  &        2.57  &        0.23   &        9.05  &        3.70  &        0.46   \\  
    5853  &       12.85  &        4.56  &        0.65   &        4.75  &        2.75  &        0.25   \\  
    5855  &        9.47  &        3.87  &        0.48   &        4.90  &        2.84  &        0.25   \\  
    5857  &       12.96  &        4.59  &        0.65   &        6.68  &        3.34  &        0.34   \\  
    5859  &       20.32  &        5.97  &        1.02   &        5.00  &        2.88  &        0.25   \\  
    5861  &       11.05  &        4.19  &        0.56   &        4.76  &        2.76  &        0.24   \\  
    5863  &       15.94  &        5.06  &        0.80   &       12.34  &        4.37  &        0.62   \\  
    5865  &        8.45  &        3.78  &        0.43   &       13.59  &        4.83  &        0.69   \\  
    5867  &       10.36  &        3.92  &        0.52   &       13.62  &        4.82  &        0.69   \\  
    5869  &        9.77  &        4.00  &        0.49   &        8.21  &        3.67  &        0.41   \\  
    5871  &       16.18  &        5.13  &        0.81   &       13.04  &        4.62  &        0.66   \\  
    5873  &       19.64  &        6.11  &        0.99   &       12.90  &        4.58  &        0.65   \\  
    5875  &        6.21  &        3.11  &        0.31   &        3.53  &        2.50  &        0.18   \\  
    5877  &        9.74  &        4.00  &        0.49   &        8.53  &        3.82  &        0.43   \\  
    5879  &        9.69  &        4.59  &        0.49   &       10.43  &        4.27  &        0.53   \\  
    5881  &       12.92  &        4.58  &        0.65   &       11.39  &        4.31  &        0.58   \\  
    5883  &       20.14  &        5.99  &        1.01   &       15.31  &        5.12  &        0.77   \\  
    5885  &       20.53  &        6.09  &        1.03   &       13.37  &        5.25  &        0.68   \\  
    5887  &       25.84  &        6.48  &        1.30   &       12.74  &        4.82  &        0.64   \\  
    5889  &       21.50  &        5.99  &        1.08   &       27.36  &        7.59  &        1.38   \\  
    5891  &       17.19  &        5.19  &        0.87   &       14.77  &        4.93  &        0.74   \\  
    5893  &       22.89  &        6.47  &        1.16   &       27.29  &        6.83  &        1.38   \\  
    5895  &       28.93  &        7.15  &        1.46   &       38.33  &        8.32  &        1.93   \\  
    5897  &       45.09  &        8.73  &        2.27   &       29.45  &        6.97  &        1.48   \\  
    5899  &       46.69  &        8.69  &        2.35   &       35.18  &        7.96  &        1.77   \\  
    5901  &       34.94  &        7.65  &        1.76   &       45.79  &        9.12  &        2.31   \\  
    5903  &       38.17  &        8.22  &        1.92   &       32.63  &        7.82  &        1.65   \\  
    5905  &       33.26  &        7.66  &        1.67   &       50.32  &        9.20  &        2.54   \\  
    5907  &       53.22  &        9.94  &        2.68   &       44.35  &        8.73  &        2.24   \\  
    5909  &       65.52  &       10.54  &        3.30   &       51.62  &        9.61  &        2.60   \\  
    5911  &       78.87  &       12.13  &        3.97   &       31.62  &        7.48  &        1.59   \\  
    5913  &       60.80  &       10.38  &        3.06   &       37.84  &        8.09  &        1.91   \\  
    5915  &       86.50  &       12.66  &        4.35   &       55.10  &       10.11  &        2.78   \\  
    5917  &       67.87  &       11.01  &        3.41   &       25.55  &        6.99  &        1.29   \\  
    5919  &       78.78  &       12.11  &        3.96   &       49.15  &        9.64  &        2.48   \\  
    5921  &       96.08  &       13.06  &        4.83   &       69.61  &       11.61  &        3.51   \\ 
\end{longtable*}

\bibliography{e10fp}
\printtables
\printfigures

\end{document}